\documentclass[journal]{IEEEtran}
\usepackage{amsmath,amsfonts}
\usepackage{algorithmic}
\usepackage{algorithm}
\usepackage{array}

\usepackage{textcomp}
\usepackage{stfloats}
\usepackage{url}
\usepackage{verbatim}
\usepackage{graphicx}
\usepackage{cite}
\usepackage{amssymb}
\usepackage{CJK}
\usepackage{indentfirst}
\usepackage{cases}
\usepackage{setspace}
\usepackage{hyperref}
\usepackage{float}
\usepackage{multirow}
\usepackage{acronym}

\acrodef{isac}[ISAC]{integrated sensing and communications}
\acrodef{sinr}[SINR]{signal-to-interference-plus-noise ratio}
\acrodef{snr}[SNR]{signal-to-noise ratio}
\acrodef{miso}[MISO]{multiple-input single-output}
\acrodef{mimo}[MIMO]{multiple-input and multiple-output}
\acrodef{mu-mimo}[MU-MIMO]{multi-user multiple-input and multiple-output}
\acrodef{mi}[MI]{mutual information}
\acrodef{crb}[CRB]{Cram{\'e}r-Rao bound}
\acrodef{dof}[DoF]{degrees of freedom}
\acrodef{sdr}[SDR]{semidefinite relaxation}
\acrodef{sdp}[SDP]{semidefinite programming}
\acrodef{moop}[MOOP]{multi-objective optimization problem}
\acrodef{svd}[SVD]{singular value decomposition}
\acrodef{music}[MUSIC]{multiple signal classification}
\acrodef{rmse}[RMSE]{root-mean-square-error}
\acrodef{mmse}[MMSE]{minimum mean square error}
\acrodef{bs}[BS]{base station}
\acrodef{csi}[CSI]{channel state information}

\ifCLASSOPTIONcompsoc
\usepackage[caption=false,font=normalsize,labelfont=sf,textfont=sf]{subfig}
\else
\usepackage[caption=false,font=footnotesize]{subfig}
\fi

\hypersetup{colorlinks=true}
\def\BibTeX{{\rm B\kern-.05em{\sc i\kern-.025em b}\kern-.08em
		T\kern-.1667em\lower.7ex\hbox{E}\kern-.125emX}}

\newtheorem{case}{\textbf{Case}}

\newtheorem{theorem}{\textbf{Theorem}}

\newtheorem{lemma}{\textbf{Lemma}}

\newtheorem{corollary}{\textbf{Corollary}}

\newtheorem{definition}{\textbf{Definition}}

\makeatletter
\newcommand{\rmnum}[1]{\romannumeral #1}
\newcommand{\Rmnum}[1]{\expandafter\@slowromancap\romannumeral #1@}
\makeatother

\usepackage{framed} 

\usepackage{cancel}

\begin{document}

\title{ Multi-Objective Optimization-based Transmit Beamforming for Multi-Target and Multi-User MIMO-ISAC Systems}
\author{
        \IEEEauthorblockN{Chunwei~Meng, Zhiqing~Wei,  Dingyou~Ma,	Wanli~Ni, Liyan~Su, and~Zhiyong~Feng}
         \thanks{This work was supported in part by the National Key Research and Development Program of China under Grant 2020YFA0711302 and 2020YFA0711303, in part by the National Natural Science Foundation of China (NSFC) under Grant U21B2014, and in part by the Beijing Municipal Natural Science Foundation under Grant L192031.}
		\thanks{C. Meng,   Z. Wei,  D. Ma and Z. Feng are 
        with the Key Laboratory of Universal Wireless Communications, Ministry of Education, Beijing University of Posts and Telecommunications,} Beijing 100876, China (e-mail: \{mengchunwei, dingyouma,  weizhiqing, fengzy\}@bupt.edu.cn).
        { Wanli Ni is with the Department of Electronic Engineering, Tsinghua University, Beijing 100084, China (e-mail: niwanli@tsinghua.edu.cn).}
        {L. Su is with the Research Department, Beijing RD Subdivision, Wireless Network, Huawei Technologies Company Ltd., Beijing  100085, China (e-mail:suliyan1@huawei.com).}
        \textit{(Corresponding author: Zhiqing~Wei,  Zhiyong~Feng)}       
    }
\maketitle
\pagestyle{plain}
\thispagestyle{plain}

\begin{abstract}
	Integrated sensing and communication (ISAC) is an enabling technology for the sixth-generation mobile communications, which equips the wireless communication networks with sensing capabilities.
      In this paper, we investigate transmit beamforming design for  multiple-input and multiple-output (MIMO)-ISAC systems in scenarios with multiple radar targets and communication users.   
      A general form of multi-target sensing mutual information (MI) is derived, along with its upper bound, which can be interpreted as the sum of individual single-target sensing MI. 
      Additionally, this upper bound can be achieved by suppressing the cross-correlation among reflected signals from different targets, which aligns with the principles of adaptive MIMO radar. %enhancing the  target detection and tracking performance.
      Then,  we propose a multi-objective optimization framework based on the signal-to-interference-plus-noise ratio of each user and the tight upper bound of sensing MI, introducing the Pareto boundary to characterize the achievable communication-sensing performance boundary of the proposed ISAC system.
      To achieve the Pareto boundary, the max-min system utility function method is employed, while considering the fairness between communication users and radar targets.
      Subsequently, the bisection search method is employed to find a specific Pareto optimal solution by solving a series of convex feasible problems.
      Finally, simulation results validate that the proposed method achieves a better tradeoff between multi-user communication and multi-target sensing performance. Additionally, utilizing the tight upper bound of sensing MI as a performance metric can enhance the multi-target resolution capability and angle estimation accuracy.

      %by utilizing a system utility function as the optimization criterion, we formulate a max-min problem to attain the Pareto boundary.
      %, which provides a flexible tradeoff between multi-target sensing and multi-user communication.
       %        Moreover, we find that the upper-bound expression exhibits a concise form similar to that of the communication sum-rate, which simplifies the problem-solving process.  
	% \textcolor{black}{       Thus, the tight upper bound of sensing MI is adopted as the sensing performance metric, subject to zero-forced inter-target interference constraints. 
       %       Then, 
    %To simultaneously optimize the multiple performance of  communication and sensing, we formulate a multi-objective optimization problem (MOOP) and transform it as a max-min utility problem to attain the Pareto boundary. 
	%	To alleviate the non-convexity of  the  problem, we convert it into a sequence of convex feasibility subproblems by incorporating slack variables.
     %     } 
	%
	%\footnote{\textcolor{black}{Rewrite according to the current version.}}.
\end{abstract}


\section{Introduction}
\IEEEPARstart{T}{he} next-generation wireless networks (6G and 5G-beyond) have been envisioned as a vital enabler for numerous emerging applications, such as autonomous vehicles, smart cities, and the Internet of Things \cite{liu2020state,liu2022integrated}.
The challenging problem is to satisfy the requirements of these applications for efficient communication and high-accuracy sensing, which motivates the development of frameworks for communication-sensing integration.
As such, \ac{isac} has been proposed as an appealing technology and has attracted great research interests recently\cite{liu2022integrated,ma2020joint,feng2021joint,li2024uav,yuan2024nest}.
In particular, \ac{isac} can significantly enhance the spectral efficiency and reduce the hardware software complexity by sharing the hardware platform and the resources in spatial, temporal as well as frequency domains for both communication and sensing \cite{zhang2021overview,zhang2022enabling,ma2020joint}.

\textcolor{black}{%Due to its spatial beamforming and waveform shaping capabilities, 
The \ac{mimo} technique plays a significant role in \ac{isac} systems, which enables spatial multiplexing, diversity, and beamforming, leading to higher data rates, improved link reliability, enhanced spatial resolution, and accurate target parameter estimation\cite{fang2023joint,wang2023nearisac}.
%However, the inherent differences between communication and sensing functionalities pose challenges in designing effective transmit beamforming strategies that can flexibly balance the performance of both functionalities while catering to diverse practical requirements.}
%
 However, the inherent difference between communication and sensing leads to compromised performance, posing challenges for \ac{mimo}-\ac{isac} systems.} % may have to change
% %
%Therefore, the key issue is to design effective beamforming strategies to achieve a favorable tradeoff in communication-sensing performance.
%
 \textcolor{black}{Therefore, the key issue is to design effective transmit beamforming strategies that can flexibly balance the performance of both functionalities while catering to the diverse requirements in practical scenarios.
Numerous studies have  investigated the joint optimization of transmit beamforming by considering both communication and sensing performance metrics \cite{liu2018mumimo,liu2022transmit,liu2020joint,chen2021joint,ni2021multi, XinyuanMI,chunwei2022,liu2021cramer,Hua2022mimo,ren2023fund,sun2023trade}.}
%For instance,\cite{liu2018mumimo} designed the transmit beamforming to approach the desired beampattern while satisfying each communication user's \ac{sinr} requirement. 
%
The work of \cite{chen2021joint} optimized the transmit beamforming by maximizing the peak sidelobe level of radar  while ensuring given \ac{sinr} threshold levels for the users.
Furthermore, the studies conducted in \cite{liu2020joint,liu2022transmit} decomposed the transmit waveform into radar and communication waveform, and optimize the beamformers for each waveform to satisfy the respective performance requirements for sensing and communication.
Addtionally, in the context of specific sensing tasks such as target estimation and tracking, some literature employs the \ac{crb} as a performance metric to evaluate the estimation performance of target parameters \cite{liu2021cramer,Hua2022mimo,ren2023fund}.
In \cite{liu2021cramer}, the authors minimized the \ac{crb} on parameter estimates for a single target while ensuring the pre-defined communication  \ac{sinr} threshold for each downlink user. 
Then, the authors in \cite{ren2023fund} minimized the multi-target estimation \ac{crb}, subject to the minimum communication requirement.
However, the lower bound for estimation accuracy provided by \ac{crb} may be not tight at low \ac{snr}, and its complex and non-convex mathematical nature always results in a more challenging optimization problem.

As a comparison, \ac{mi} is also a commonly used 
 sensing performance metric, which provides accurate estimation and classification capabilities in a more concise form\cite{bellmi,zhang2022enabling}.
In \cite{tang2018spectrally}, the authors demonstrated that the optimal waveform designed by maximizing sensing \ac{mi} enables efficient coexistence of \ac{mimo} radar and communication systems occupying the same spectrum.
In \cite{yang2007mimo}, the authors showed that under the assumption of Gaussian distributed target response, maximizing \ac{mi} and maximizing \ac{mmse} lead to the same optimal waveform solution.
In \cite{ni2021multi},  the emphasis on the accuracy of entire sensing channel vectors with sensing \ac{mi} enhances overall sensing performance beyond the focus on specific parameter-associated partial channels with the \ac{crb}.
The authors in \cite{chen2013adapt} showed that  maximizing sensing \ac{mi} through the waveform design improved the target detection and feature extraction performance. 
\textcolor{black}{Besides, the similarity between the communication and sensing \ac{mi} expressions also facilitates the efficient tradeoff between the two functionalities through weighted sum optimization\cite{XinyuanMI,wei2024wave}.}
\textcolor{black}{The authors in \cite{XinyuanMI} optimized a weighted sum of communication and sensing \ac{mi} to improve the balanced performance of both functionalities. }
In \cite{dong2023rethink}, the authors established a relationship between sensing \ac{mi} and the rate-distortion theory, imparting operational estimation theoretic meaning to \ac{mi}-based methods.
Furthermore, \cite{li2023mutual} demonstrated that \ac{mi}-based beamforming design can effectively suppress echo interference from  scatters in the surrounding environment.
%
%

%The authors in  optimized a weighted sum of communication and sensing \ac{mi} to improve the balanced performance of both functionalities. 
%

Simultaneously supporting multi-user communication and multi-target sensing in practical scenarios poses a critical challenge for \ac{mimo}-\ac{isac} systems\cite{ren2023fund,chen2021joint}.
\textcolor{black}{Such scenarios are inherently complex due to the diverse performance requirements for both communication and sensing. 
Moreover, they involve multiple intrinsic tradeoffs, such as the performance tradeoffs among multiple communication users, multiple sensing targets, and between communication and sensing.
Therefore, it is essential to develop a beamforming method that can flexibly balance multi-user communication and multi-target sensing performance based on their respective  priorities.
Given its advantages in enhancing target detection, estimation, and classification performance in a concise manner, sensing \ac{mi} serves as a more suitable performance metric for multi-target sensing, as mentioned before\cite{chen2013adapt, ni2021multi,wei2024wave, dong2023rethink, li2023mutual, XinyuanMI,chunwei2022}.
However, existing research based on sensing \ac{mi} still has some limitations.}
% due to their inherent complexity and varied performance requirements. 
%Because such scenarios are intrinsically complex and entail diverse performance requirements for both communication and sensing.
%
%Moreover, task-specific performance metrics, like detection probability and the \ac{crb}, may not effectively enhance overall sensing performance in diverse multi-target scenarios.
%
%
%
Firstly, sensing \ac{mi} in current studies offers a general overview of the sensing channel containing the multi-target information, but lacks a clear and detailed characterization of multi-target sensing performance and the relationships between targets.
Secondly, the lack of precise depiction of achievable performance boundaries for communication and sensing hinders the attainment of optimal performance tradeoffs.
Furthermore, the absence of fairness consideration between users and targets poses limitations on meeting diverse and specific requirements, thereby impeding the flexibility of beamforming designs.

\textcolor{black}{To address these limitations, we  propose a novel transmit beamforming approach for multi-user multi-target \ac{mimo}-\ac{isac} systems  based on the sensing \ac{mi} and communication \ac{sinr}s. }%facilitating  multi-target sensing and multi-user communication for \ac{mimo}-\ac{isac} systems. 
We consider a scenario where the \ac{bs} communicates with multiple downlink users while sensing multiple targets simultaneously.
To fully exploit the \ac{dof}s provided by \ac{mimo} to meet the performance requirements of both multi-user communication and multi-target sensing, we synthesize communication and radar signals, and then jointly design  their respective transmit beamforming\cite{liu2020joint,liu2022transmit,liu2021cramer}.
%Due to the mentioned benefits of sensing \ac{mi} and its concise and tractable expression, we adopt it as the metric to measure the performance of multi-target sensing.
%We then derive a general expression for sensing \ac{mi}, which is found to depend on the cross-correlation between reflected signals from multiple targets and the statistical properties of each target's complex reflection coefficient.
%
Then, we derive a general expression for sensing \ac{mi} and its tight upper bound to explore the structural features of maximum sensing \ac{mi}.
We find that maximizing the \ac{mi} upper bound under zero-forced cross-correlation constraints aligns with the principles of adaptive \ac{mimo} radar technique\cite{lijian2007}.
Therefore, we utilize the constrained upper bound of sensing \ac{mi} as the performance metric for multi-target sensing, each user's \ac{sinr} as the communication performance metric, and construct a \ac{moop} to comprehensively investigate the tradeoff in communication-sensing performance.
To efficiently solve the \ac{moop}, we define the achievable performance region and its Pareto boundary. 
Then, we formulate a max-min utility optimization problem to obtain a specific Pareto solution for the \ac{moop}, 
and utilize the properties of the achievable performance region to prove that the optimal solution of this max-min problem lies on the Pareto boundary.
Finally,  a  bisection search method is employed to find the Pareto optimal solution for  a specific set of communication-sensing weights. 
The main contributions of this paper are summarized as follows:
\begin{itemize}
		\item % 
       We derive a novel general form of multi-target sensing \ac{mi} and alongside a tight upper bound that satisfies zero-forced cross-correlation constraints.
       Furthermore, by suppressing the cross-correlation among signals reflected from different targets, not only can sensing \ac{mi} reach its maximum, but the signals can also be considered independent of each other, which benefits target detection and tracking.
       Remarkably, we find that the upper bound of sensing \ac{mi} can be viewed as the sum of single-target sensing \ac{mi}, resembling a concise form similar to communication sum-rate and simplifying subsequent optimization processes.
        
	  \item  %
      To comprehensively investigate the tradeoff in communication and sensing performance, we formulate a \ac{moop} to simultaneously optimize communication \ac{sinr}s and sensing \ac{mi}.
      Then, we  introduce the Pareto boundary of the \ac{moop} to characterize the achievable performance boundary of the proposed \ac{isac} system.
      To obtain a specific set of Pareto solutions, a max-min  utility function method is employed, which can be solved through a bisection search algorithm.
      %to transform the problem into starting from an initial point determined by performance thresholds and searching along the weight direction for points that intersect with the Pareto boundary.
      %, 
      This method provides a flexible tradeoff between multi-target sensing and multi-user communication performance, while meeting specific sensing and communication requirements and their respective priorities.
       
	\item  %
      Finally, numerical simulations demonstrate the following: i) The necessity of transmitting additional radar signals to provide sufficient \ac{dof}s for effectively resolving multiple targets, particularly when the number of targets surpasses the number of users.
      ii)  The adoption of the constrained sensing \ac{mi} upper bound as the performance metric for multi-target sensing offers an enhanced tradeoff between sensing and communication performance. 
      Moreover, it also significantly improves resolution for closely located multiple targets and enhances accuracy in angle estimation.      
\end{itemize}

% We first introduce a dedicated radar signal to the conventional communication signal, and then jointly design the transmit beamforming for both sensing and communication, which guarantees the sensing accuracy even for a relatively large number of targets.
% %
% However, the multi-target sensing \ac{mi}  derived from the synthesized signal is intrinsically complex and non-convex. 
% %
% To simplify the problem, we derive the tight upper bound of sensing \ac{mi}, which is proved to be a function related to the \ac{sinr}s of radar targets, and have a concise  structure similar to that of the communication sum-rate. 
% %
% Since the tight upper bound simplifies the subsequent optimization procedures, we use it as  the sensing performance metric for the \ac{isac} system.
% %
% To simultaneously optimize the multi-user communication and multi-taget sensing performance, we formulate a \ac{moop} based on the \ac{sinr} of each user and the sensing \ac{mi}.
% %
% To efficiently solve the \ac{moop}, we define the achievable performance region and its Pareto boundary. %based on the  \ac{sinr} of each user and the tight upper bound of sensing \ac{mi}. 
% 
%

The rest of this paper is organized as follows.
Section~\ref{section_system_model} introduces the system model along with the performance metric for multi-user communications. 
Section~\ref{section_sensingMI} derives the sensing \ac{mi} and its tight upper bound for multi-target  sensing. 
Section~\ref{section_optimization} investigates the multi-objective optimization for \ac{isac} beamforming, where we employ a max-min utility function method to solve the \ac{moop}. 
Simulation results are presented in Section~\ref{section_simulation}.
Finally, we provide concluding remarks in Section~\ref{section_conclusion}.

\textit{Notations}:
Boldface lowercase and uppercase letters  denote vectors and matrices, respectively.
The set of complex number is  $\mathbb{C}$.
The transpose, conjugate
transpose, conjugate, inverse and pseudo-inverse operation is denoted by  $(\cdot)^T$, $(\cdot)^H$, $(\cdot)^{*}$ , $(\cdot)^{-1}$ and $(\cdot)^{\dagger}$,  respectively. The  expected value of a random argument is denoted by $\mathbb{E}(\cdot)$. 
We let $\otimes$ denote the Kronecker product, and let  $\mathbf{I}_M$ denote the $M$-dimensional identity matrix.
The curled inequality symbol $\succeq$ is utilized to denote the generalized matrix inequality, i.e., $\mathbf{A} \succeq \mathbf{0}$ means that $\mathbf{A}$ is positive semi-definite matrix.
The symbols $\text{det}(\cdot)$ and $\text{tr}(\cdot)$ denote the determinant and trace of a matrix, respectively.

\section{System Model} \label{section_system_model}
\begin{figure} [t]
	\centering
	\includegraphics[width=0.45 \textwidth]{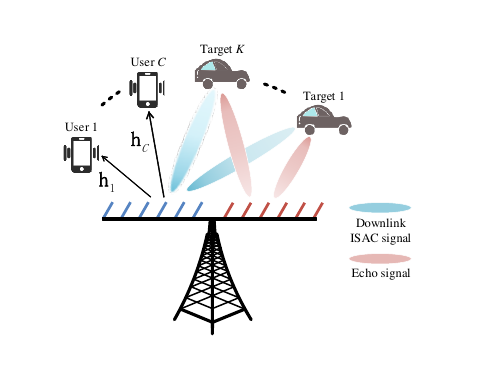}
	\caption{Illustration of the \ac{isac}  scenario  where the BS serves $C$ downlink communication users while detecting $K$ targets.}
	\label{system_model}
\end{figure}

We consider a colocated \ac{isac} system as shown in Fig.~\ref{system_model}, where the BS is equipped with ${{N}_{t}}$ transmit antennas and ${N}_{r}$ receive antennas as uniform linear arrays (ULAs). 
%[JSAC-2022] Joint
The BS aims to serve $C$ downlink single-antenna users while detecting $K$ targets. % [TSSP-2020] \ac
The set of communication users is indexed by $\mathcal{C} = \{ 1,2, \ldots, C\}$. 
To achieve the satisfactory performance of  communication and radar sensing, we exploit the maximum \ac{dof}s provided by \ac{mimo} and transmit  ${M}$ additional probing streams.
The transmit signal is a sum of precoded communication signals and radar probing signals \cite{liu2020joint,liu2021cramer,liu2022transmit}, i.e.,
\begin{equation}
	\label{eq_X}
	\mathbf{X}={{\mathbf{W}}_{c}}{{\mathbf{S}}_{c}}+{{\mathbf{W}}_{r}}{{\mathbf{S}}_{r}},
\end{equation}
% [JSAC-2022] xiangliu
where $\mathbf{X}\in\mathbb{C}^{N_{t} \times L}$ is the transmit signal matrix with $L>{N}_{t}$ being the length of the signals.
The $i$th data stream of the communication is  denoted by ${\mathbf{s}}_{i}$, $\forall i \in \mathcal{C}$, and ${{\mathbf{S}}_{c}}={{\left[ \mathbf{s}_{1}^{T},\ldots ,\mathbf{s}_{C}^{T} \right]}^{T}}\in {{\mathbb{C}}^{{{C}}\times {{L}}}}$ contains $C$ data streams intended for the ${C}$ users.
Similarly,  the $m$-th probing stream of radar sensing is  denoted by   ${\mathbf{s}}_{C+m}$, $m\in\left\{1, 2, \cdots, M\right\}$, and ${{\mathbf{S}}_{r}}={{\left[ \mathbf{s}_{{C}+1}^{T},\ldots ,\mathbf{s}_{{C}+M}^{T} \right]}^{T}}\in {{\mathbb{C}}^{{{M}}\times {{L}}}}$ contains $M$ individual probing streams with $M < N_{t}-C$\cite{liu2021cramer}.
The matrices ${{\mathbf{W}}_{c}}=\left[ {{\mathbf{w}}_{1}}, \ldots,  {{\mathbf{w}}_{{{C}}}} \right]\in {{\mathbb{C}}^{{{N}_{t}}\times {{C}}}}$ and ${{\mathbf{W}}_{r}}=\left[ {{\mathbf{w}}_{{C}+1}},\ldots, {{\mathbf{w}}_{{C}+M}} \right]\in {{\mathbb{C}}^{{{N}_{t}}\times {M}}}$ contain the transmit beamforming vectors for the data streams and the probing streams, respectively.

We assume that both the communication signals and radar probing signals are  wide-sense stationary stochastic processes with zero mean and unit power\cite{liu2020joint}.
The communication data signals for different users are uncorrelated, so $\frac{1}{L}\mathbb{E}\left[{{\mathbf{S}}_{c}}\mathbf{S}_{c}^{H}\right]={{\mathbf{I}}_{{{C}}}}$.
The radar probing signals are pseudorandom sequences with zero mean and unit variance, and are uncorrelated with each other\cite{Sarwate1980cross}, i.e., $\frac{1}{L}\mathbb{E}\left[{{\mathbf{S}}_{r}}\mathbf{S}_{r}^{H}\right]={{\mathbf{I}}_{{{M}}}}$.
The communication signals and radar probing signals are assumed to be uncorrelated, namely  $\mathbb{E}\left[{{\mathbf{S}}_{c}}\mathbf{S}_{r}^{H}\right]={{\mathbf{0}}}$.
Then, we can derive the covariance matrix of the transmit signal, given by  
\begin{equation}
	\label{R_X}
	\mathbf{R}_{X}=\frac{1}{L}\mathbb{E}\left[ \mathbf{X}{{\mathbf{X}}^{H}} \right]=\sum\limits_{n=1}^{C+M}{{{\mathbf{w}}_{n}}\mathbf{w}_{n}^{H}},%={{\mathbf{W}}_{r}}\mathbf{W}_{r}^{H}+{{\mathbf{W}}_{c}}\mathbf{W}_{c}^{H}.
\end{equation}

In the following, we describe the received signal model and the performance metric for multi-user communications in Subsection~\ref{section_MUcom_model}.
Subsequently, we present the receive signal model for radar sensing at the BS in Subsection~\ref{section_radar_model}.

\subsection{Multi-User Communication Model}
\label{section_MUcom_model}
For downlink communications, the signal received at the $i$-th user is expressed as
\begin{equation}
	\label{rs_com}
	\begin{aligned}
		{{\mathbf{y}}_{i}} & =\mathbf{h}_{i}^{H}\mathbf{X}+{{\mathbf{z}}_{i}},\\
	\end{aligned}
\end{equation}
where ${{\mathbf{h}}_{i}}\in {{\mathbb{C}}^{{{N}_{t}}\times 1}}$ is the channel matrix between the BS and the $i$-th user, 
and ${{\mathbf{z}}_{i}}\sim\mathcal{C}\mathcal{N}( 0,{\sigma_{i}^{2}\mathbf{I}_{L}})$ is a complex additive white Gaussian noise (AWGN) vector with zero mean and covariance $\sigma_{i}^{2}\mathbf{I}_{L}$.

Since the communication users have no prior information about the probing streams, the users suffer from the interference caused by the probing streams and the multi-user interference\cite{liu2020joint,liu2021cramer }.
Specifically, the received signal (\ref{rs_com}) can be rewritten as
\begin{equation}
	\label{rs_com2}
	\begin{aligned}
		{{\mathbf{y}}_{i}} = \mathbf{h}_{i}^{H}{{\mathbf{w}}_{i}}{{\mathbf{s}}_{i}}+\!\!\! \sum\limits_{j=1,j\ne i}^{C}{\mathbf{h}_{i}^{H}{{\mathbf{w}}_{j}}}{{\mathbf{s}}_{j}}+\!\!\!\sum\limits_{j=C+1}^{C+M}{\mathbf{h}_{i}^{H}{{\mathbf{w}}_{j}}}{{\mathbf{s}}_{j}} + {{\mathbf{z}}_{i}} ,  \forall i \in \mathcal{C},
	\end{aligned}
\end{equation}
where  the initial term is the useful signal, while the second term denotes the interference caused by the other communication users. The third term indicates the interference originating from the radar probing signals.
Thus, the  \ac{sinr} of the $i$-th user is 
\begin{equation}
	\label{sinr_com}
	\begin{aligned}
		{{\gamma }_{i}}=\frac{{{\left| \mathbf{h}_{i}^{H}{{\mathbf{w}}_{i}} \right|}^{2}}}{\sum_{j=1,j\ne i}^{C}{{{\left| \mathbf{h}_{i}^{H}{{\mathbf{w}}_{j}} \right|}^{2}}}+\sum_{j=C+1}^{C+M}{{{\left| \mathbf{h}_{i}^{H}{{\mathbf{w}}_{j}} \right|}^{2}}}+\sigma _{i}^{2}}.	
	\end{aligned}
\end{equation}

The overall performance of the \ac{mu-mimo} communication is evaluated by the average rate, which is given by \cite{robust2013fritz}
\begin{equation}
	\label{averate_com}
	\begin{aligned}
		{{r}_{c}}=\frac{1}{C}\left( \sum\limits_{i=1}^{C}{\log \left( 1+{{\gamma }_{i}} \right)} \right).	
	\end{aligned}
\end{equation}

\subsection{Radar Sensing Model}
\label{section_radar_model}
The BS uses the reflected echoes to recover the parameters of the targets.
In this paper, we focus on beamforming in the spatial domain. For brevity, the sensing targets are modeled to be stationary, and are located at the same range resolution as assumed in \cite{liu2018mumimo, liu2018toward, liu2021cramer}. With these assumptions, the target response matrix is expressed as the superposition of the response of the individual target~\cite{Moura2008time}, i.e.,
\begin{equation}
	\label{sen_channel}
	\mathbf{G}=\sum\limits_{k=1}^{K}{{{\beta }_{k}}\text{ }\mathbf{b}\left( {{\theta }_{k}} \right)}\text{ }{{\mathbf{a}}^{H}}\left( {{\theta }_{k}} \right),
\end{equation}
where $\mathbf{a}\left( \theta_{k}  \right)={{[ 1,{{e}^{j\frac{2\pi }{\lambda }{d}\sin \left( \theta_{k}  \right)}},\ldots, {{e}^{j\frac{2\pi }{\lambda }\left( {{N}_{t}}-1 \right){d}\sin \left( \theta_{k}  \right)}} ]}^{T}}\in {{\mathbb{C}}^{{{N}_{t}}\times 1}}$ and $\mathbf{b}\left( \theta_{k}  \right)=[ 1,{{e}^{j\frac{2\pi }{\lambda }{d}\sin \left( \theta_{k}  \right)}},\ldots,$ ${{{e}^{j\frac{2\pi }{\lambda }\left( {{N}_{r}}-1 \right){d}\sin \left( \theta_{k}  \right)}} ]}^{T}\in {{\mathbb{C}}^{{{N}_{r}}\times 1}}$ are the corresponding transmit and receive steering vectors of the echo with the direction at $\theta_{k}$, respectively.
In general, the target can be modeled as being composed of an infinite number of random, isotropic and independent scatterers over the area of interest, and the complex gain of each scatterer can be modeled as a zero-mean and white complex random variable \cite{2018cheng}.
 Together with the fact that the incident angles between different targets are independently distributed \cite{lehmann2007evaluation}, the complex coefficients of different targets can be assumed to be independently Gaussian distributed, i.e., $\beta_k \sim \mathcal{C N}\left(0, {\sigma}_k^2\right), \forall k$ \cite{hua2023joint}.

The distance of adjacent antenna elements is denoted by  $d$, and  $\lambda$ denotes the wavelength.
Thus, the signal received  at the BS is given by
\begin{equation}
	\label{received_signal_sensing}
	{{\mathbf{Y}}_{r}}=\mathbf{GX}+{{\mathbf{Z}}_{r}},
\end{equation}
where ${{\mathbf{Z}}_{r}}\in {{\mathbb{C}}^{{{N}_{r}}\times L}}$ is a complex  AWGN matrix with zero mean and covariance  $\sigma _{r}^{2}\mathbf{I}_{L}$. 

Upon vectorizing (\ref{received_signal_sensing}), the received signal is recast as
\begin{equation}
	\label{received_signal_sen_vec}
	{{\mathbf{\tilde{y}}}_{r}}=\mathbf{\tilde{X}\tilde{g}}+{{\mathbf{\tilde{z}}}_{r}},
\end{equation}
where ${{\mathbf{\tilde{y}}}_{r}}\triangleq\text{vec}( {{\mathbf{Y}}_{r}} )$, $\mathbf{\tilde{X}}\triangleq( {{\mathbf{X}}^{T}}\otimes {{\mathbf{I}}_{{{N}_{r}}}})$, $\mathbf{\tilde{g}}\triangleq\text{vec}( \mathbf{G} )$, and ${{\mathbf{\tilde{z}}}_{r}}\triangleq\text{vec}( {{\mathbf{Z}}_{r}} )$. 
Let $ {\mathbf{R}}_{G}
=\mathbb{E}[ \mathbf{\tilde{g}}{{{\mathbf{\tilde{g}}}}^{H}}]$ be the spatial correlation matrix of $\mathbf{G}$, and  $ {\mathbf{R}}_{Z}
=\mathbb{E}[ {{\mathbf{\tilde{z}}}_{r}}{{{\mathbf{\tilde{z}}}_{r}}^{H}} ]$ be the covariance matrix of ${{\mathbf{\tilde{z}}}_{r}}$.
To facilitate the derivation, we assume $\mathbf{\tilde{g}}\sim\mathcal{C}\mathcal{N}( \mathbf{0},{{\mathbf{R}}_{G}} )$, which is consistent with the classic literature on radar \ac{mi}\cite{bellmi,liuAdaptive,tang2018spectrally}.
Then, we have ${{\mathbf{\tilde{z}}}_{r}}\sim\mathcal{C}\mathcal{N}( \mathbf{0},{{\mathbf{R}}_{Z}} )$ and ${{\mathbf{\tilde{y}}}_{r}}\sim\mathcal{C}\mathcal{N}( \mathbf{0},\mathbf{\tilde{X}}{{\mathbf{R}}_{G}}{{{\mathbf{\tilde{X}}}}^{H}}+{{\mathbf{R}}_{Z}})$. 
The covariance matrix ${\mathbf{R}}_{G}$ is given by  
\begin{equation}
	\label{r_g}
			{\mathbf{R}}_{G}
			 \!=\!\mathbb{E}\left[ \mathbf{\tilde{g}}{{{\mathbf{\tilde{g}}}}^{H}} \right]
			\!\overset{\left(a\right)}{=}\!\sum_{k=1}^{K}{\sigma _{k}^{2}\left({{\mathbf{a}}^{*}}\!\!\left( {{\theta }_{k}} \right)\! \otimes \! \mathbf{b}\left( {{\theta }_{k}} \right) \right){{\left( {{\mathbf{a}}^{*}}\!\!\left( {{\theta }_{k}} \right)\! \otimes  \!\mathbf{b}\left( {{\theta }_{k}} \right) \right)}^{H}}},
\end{equation}
where $\left(a\right)$  follows $\text{vec}(\mathbf{\mathbf{b}\mathbf{a}^{H}})=({{\mathbf{a}}^{*}}\otimes \mathbf{b})$.

To demonstrate the necessity of  transmitting additional radar signals, we consider a scenario where only $C$ communication signals are transmitted, i.e., $\mathbf{X}={{\mathbf{W}}_{c}}{{\mathbf{S}}_{c}}$.
In this case, the number of  \ac{dof}s for transmit bemaforming design is limited by the number of communication users due to the fact that $\text{rank}(\mathbf{R}_X)=C$.
However, when there exists $K$ targets with $K > C$, the number of available \ac{dof}s is insufficient to for accurate estimation of the targets using angle estimation techniques such as \ac{music} and Capon\cite{lijian2007}.
Consequently, this leads to an inevitable degradation in the performance of multi-target sensing due to  the lack of radar \ac{dof}s\cite{liu2021cramer}. 
Therefore, it is imperative to transmit additional radar signals in scenarios with a large number of targets.
The detailed analysis is provided in Section~\ref{section_simulation}.

\section{Mutual Information for Radar Sensing} \label{section_sensingMI}
In order to characterize the  performance of multi-target sensing, we introduce the sensing \ac{mi} which can be employed to measure how much environmental information can be observed in the BS\cite{bellmi}.
The sensing \ac{mi} is generally defined as the conditional \ac{mi} between the sensing channel $\mathbf{\tilde{g}}$ and the received signal ${{{\mathbf{\tilde{y}}}}_{r}}$ with the given transmit signal\cite{XinyuanMI}.
Following \cite{tang2010mimo}, we obtain the general expression of sensing \ac{mi} as
\begin{equation}
	\begin{aligned}
	I\left( {{{\mathbf{\tilde{y}}}}_{r}};\mathbf{\tilde{g}}\left| {\mathbf{\tilde{X}}} \right. \right)	%
	& =\log \left[ \det\! \left( \mathbf{\tilde{X}}{{\mathbf{R}}_{G}}{{{\mathbf{\tilde{X}}}}^{H}}+{{\mathbf{R}}_{Z}} \right) \right]\!-\!\log \left[ \det \left( {{\mathbf{R}}_{Z}} \right) \right] \\
		& =\log \left[ \det \left( \mathbf{I}+\mathbf{\tilde{X}}{{\mathbf{R}}_{G}}{{{\mathbf{\tilde{X}}}}^{H}}\mathbf{R}_{Z}^{-1} \right) \right] \label{MI_1} \\ 
		& \overset{\left(a\right)}{=}\log \left[ \det \left( \mathbf{I}+\frac{1}{\sigma _{r}^{2}}{{\mathbf{R}}_{G}}{{{\mathbf{\tilde{X}}}}^{H}}\mathbf{\tilde{X}} \right) \right],	
	\end{aligned}
\end{equation}
where $\mathbf{R}_{z}$ is the covariance matrix of ${{{\mathbf{\tilde{z}}}}_{r}}$, and $\left(a\right)$ follows  Sylvester's determinant theorem, i.e., $\det \left( \mathbf{I}+\mathbf{AB} \right)=\det \left( \mathbf{I}+\mathbf{BA}\right)$.
Due to the fact that the columns of the noise 
${{\mathbf{Z}}_{r}}$ are independent of each other, we have ${{\mathbf{R}}_{Z}}={\sigma _{r}^{2}}{{\mathbf{I}}_{L{{N}_{r}}}}$.
 
Substituting the specific expression of $\mathbf{\tilde{X}}$ into (\ref{MI_1}), the sensing \ac{mi} can be rewritten as 
\begin{align}
		\label{sen_mi_3_22}
	I\left( {{{\mathbf{\tilde{y}}}}_{r}};\mathbf{\tilde{g}}\left| {\mathbf{\tilde{X}}} \right. \right) & =\! \log \! \left[ \det \! \left(\!  \mathbf{I}\! +\! \frac{1}{\sigma _{r}^{2}}{{\mathbf{R}}_{G}}{{\left( {{\mathbf{X}}^{T}}\! \otimes {{\mathbf{I}}_{{{N}_{r}}}} \! \right)} ^{H}}\! \! \left( {{\mathbf{X}}^{T}}\! \otimes {{\mathbf{I}}_{{{N}_{r}}}} \! \right)\!  \right)\!  \right] \notag \\ 
    & \overset{\left(a\right)}{=}\! \log \left[ \det \left( \mathbf{I}+\frac{1}{\sigma _{r}^{2}}\left( {{\mathbf{X}}^{*}}{{\mathbf{X}}^{T}}\otimes {{\mathbf{I}}_{{{N}_{r}}}} \right){{\mathbf{R}}_{G}} \right) \right],  
\end{align}
where $\left(a\right)$ is based on the property of the Kronecker product  $(\mathbf{AB})\otimes (\mathbf{CD})=(\mathbf{A}\otimes \mathbf{C})(\mathbf{B}\otimes \mathbf{D})$ and $\det \left( \mathbf{I}+\mathbf{AB} \right)=\det \left( \mathbf{I}+\mathbf{BA}\right)$. 
However, the expression of sensing \ac{mi} in (\ref{sen_mi_3_22}) is intractable due to the Kronecker product structure. 
Therefore, we  recast the sensing \ac{mi} into a more concise and intuitive form in Lemma \ref{lemma_mi_trans}.
\begin{lemma}\label{lemma_mi_trans}
    \emph{ By performing mathematical transformations, we obtain a novel form of sensing \ac{mi}, namely,
    \begin{equation}
	\label{sen_mi_phi}
	\begin{aligned}
		I\left( {{{\mathbf{\tilde{y}}}}_{r}};\mathbf{\tilde{g}}\left| {\mathbf{\tilde{X}}} \right. \right) &=\log \left[  \det \left( \mathbf{\Phi }+\mathbf{\Lambda } \right)  \prod \limits_{k=1}^K\sigma _{k}^{2}\right],\\
	\end{aligned}
\end{equation}
where the positive semi-definite matrix $\mathbf{\Phi }\in {{\mathbb{C}}^{K\times K}}$ and diagonal matrix $\mathbf{\Lambda }\in {{\mathbb{C}}^{K\times K}}$ are respectively defined by
\begin{equation}
	\label{phi}
	\begin{aligned}
		 {{\left[ \mathbf{\Phi } \right]}_{i,j}}&\triangleq \frac{L}{\sigma _{r}^{2}}\text{tr}\left( {{\mathbf{A}}^{H}}\left( {{\theta }_{i}} \right)\mathbf{A}\left( {{\theta }_{j}} \right){{\mathbf{R}}_{X}} \right) \\
		&= \frac{{\alpha }_{ij}L}{\sigma _{r}^{2}}{{\mathbf{a}}^{H}}\left( {{\theta }_{i}} \right)\left( \sum\nolimits_{n=1}^{{{C}}+M}{{{\mathbf{w}}_{n}}\mathbf{w}_{n}^{H}}\right)\mathbf{a}\left( {{\theta }_{j}} \right),
	\end{aligned}
\end{equation}
and
\begin{equation}
	\label{phi2}
	\begin{aligned}
		 \mathbf{\Lambda}\triangleq \text{diag}\left\{ \frac{1}{\sigma _{1}^{2}},\frac{1}{\sigma _{2}^{2}},\ldots ,\frac{1}{\sigma _{K}^{2}} \right\},
	\end{aligned}
\end{equation}
where ${{\left[ \mathbf{\Phi } \right]}_{i,j}}$ is the $(i,j)$-th element of $\mathbf{\Phi }$, $\mathbf{A}\left( {{\theta }_{k}} \right)\triangleq \mathbf{b}\left( {{\theta }_{k}} \right){{\mathbf{a}}^{H}}\left( {{\theta }_{k}} \right)$, and ${{\alpha }_{ij}}\triangleq{{\mathbf{b}}^{H}}\left( {{\theta }_{i}} \right)\mathbf{b}\left( {{\theta }_{j}} \right)$.    }
\end{lemma}
\begin{IEEEproof}
	Please refer to Appendix \ref{app_mi}.
\end{IEEEproof}

It is noted that the elements of $\mathbf{\Phi }$ in (\ref{phi}) are the cross-correlation between the signals reflected back from the targets of interest, defined as the cross-correlation pattern\cite{Stoica2007}.
Therefore, (\ref{sen_mi_phi}) indicates that the multi-target sensing \ac{mi} relies on the statistical characteristics of the reflection coefficients $\beta_k, \forall k$ and cross-correlation pattern among signals.
However, (\ref{sen_mi_phi}) is still complex and intractable due to the coupling of each element in $\mathbf{\Phi }$ via the beamforming vectors.
To simplify the expression and obtain the structural features of the maximum sensing \ac{mi}, we derive an upper bound of sensing \ac{mi} as shown below.
\begin{theorem}\label{lemma_mi}
	\emph{The upper bound of sensing \ac{mi} is denoted by $I_{\rm up}$, and we have
 \begin{equation}
	\label{upb_senmi_1}
	\begin{aligned}
		{{I}_{\rm up}}=\!\sum\limits_{\text{k}=1}^{K}{\log\! \left(1+ \!\frac{{{N}_{r}}\sigma _{k}^{2}L}{\sigma _{r}^{2}}{{\mathbf{a}}^{H}}\!\!\left( {{\theta }_{k}} \right)\!\left(\! \sum\limits_{n=1}^{C+M}{{{\mathbf{w}}_{n}}\mathbf{w}_{n}^{H}\!} \right)\!\mathbf{a}\left( {{\theta }_{k}} \right) \right)}.
	\end{aligned}
\end{equation}
     The bound is tight when
     \begin{equation}
	\label{upb_const}
	\begin{aligned}
		{{\mathbf{a}}^{H}}\left( {{\theta }_{j}} \right)\!\left( \!\sum\nolimits_{n=1}^{{{C}}+M}{{{\mathbf{w}}_{n}}\mathbf{w}_{n}^{H}}\!\right)\!\mathbf{a}\left( {{\theta }_{i}} \right)=0, \forall i,j \left(i\ne j \right),
	\end{aligned}
\end{equation} }
where ${{\alpha }_{ji}}={{\mathbf{b}}^{H}}\left( {{\theta }_{j}} \right)\mathbf{b}\left( {{\theta }_{i}} \right)$.
\end{theorem}

\begin{IEEEproof}
Based on Lemma \ref{lemma_mi_trans}, applying the  Hadamard's inequality\cite[Sec. 6.1]{meyer2000matrix} for the determinant of an ${N\times N}$ positive semi-definite matrix to  (\ref{sen_mi_phi}), we obtain that 
\begin{equation}
	\label{hardamad}
	\begin{aligned}
		&\log \left[ \det \left( \mathbf{\Phi }+\mathbf{\Lambda } \right) \prod \limits_{k=1}^K\sigma _{k}^{2} \right]\le \\ & \log \left[ \prod\limits_{i=1}^{K}{\left(1+ \frac{{{\alpha }_{ii}}\sigma _{i}^{2}L}{\sigma _{r}^{2}}{{\mathbf{a}}^{H}}\left( {{\theta }_{i}} \right)\left( \sum\nolimits_{n=1}^{{{C}}+M}{{{\mathbf{w}}_{n}}\mathbf{w}_{n}^{H}}\right)\mathbf{a}\left( {{\theta }_{i}} \right)\right)} \right],
	\end{aligned}
\end{equation}
with equality holding if and only if $\mathbf{\Phi }+\mathbf{\Lambda }$ is diagonal, i.e.,
\begin{equation}
	\label{upb_const11}
	\begin{aligned}
		{{\mathbf{a}}^{H}}\left( {{\theta }_{j}} \right)\left( \sum\nolimits_{n=1}^{{{C}}+M}{{{\mathbf{w}}_{n}}\mathbf{w}_{n}^{H}}\right)\mathbf{a}\left( {{\theta }_{i}} \right)=0\text{    }\left(i\ne j \right).
	\end{aligned}
\end{equation}
 Substituting $\alpha_{i,i} = N_r$ into the right-hand side of (\ref{hardamad}), we can obtain the upper bound of sensing \ac{mi} in (\ref{upb_senmi_1}).
\end{IEEEproof}

The equation \eqref{upb_const} represents the zero-forced cross-correlation constraints, indicating that the signals reflected from different targets are independent of each other. 
When these constraints are satisfied, the upper bound of sensing \ac{mi} can be achieved. 
However, fully eliminating the cross-correlation pattern is highly challenging.  In practical applications, it is often approximated by imposing constraints to ensure that the  absolute value  of cross-correlation pattern is smaller than a sufficiently low threshold value\cite{hua2023joint}, namely,
\begin{equation} \label{appro_zf_cons}
    \begin{aligned}
        \left|{{\mathbf{a}}^{H}}\left( {{\theta }_{j}} \right)\left( \sum\nolimits_{n=1}^{{{C}}+M}{{{\mathbf{w}}_{n}}\mathbf{w}_{n}^{H}}\right)\mathbf{a}\left( {{\theta }_{i}} \right)\right| \le \kappa, \left(i\ne j \right),
    \end{aligned}
\end{equation}
where $\kappa \to 0$.

Additionally, we see that the second term  in (\ref{upb_senmi_1}) is the  \ac{sinr} of the reflected echo from the $k$-th target, which is denoted by 
\begin{equation}
	\label{upb_sinr_k}
	\begin{aligned}
		{\text{SINR}_{k}}=\frac{{{N}_{r}}\sigma _{k}^{2}L}{\sigma _{r}^{2}}{{\mathbf{a}}^{H}}\!\!\left( {{\theta }_{k}} \right)\!\left(\! \sum\limits_{n=1}^{C+M}{{{\mathbf{w}}_{n}}\mathbf{w}_{n}^{H}\!} \right)\!\mathbf{a}\left( {{\theta }_{k}} \right).
	\end{aligned}
\end{equation}
By substituting (\ref{upb_sinr_k}) into (\ref{upb_senmi_1}), we have the following insight given in Corollary 1.
\begin{corollary}
	\emph{The upper bound of sensing \ac{mi} can be expressed in a form similar to the communication sum-rate, i.e.,}
\begin{equation}
	\label{upb_senmi_sinr}
	\begin{aligned}
		{{I}_{\rm up}}=\sum\nolimits_{\text{k}=1}^{K}{\log \left( 1+{\text{SINR}_{k}} \right)},
	\end{aligned}
\end{equation}
 \emph{  where the component $\log \left( 1+{\text{SINR}_{k}} \right)$ in ${{I}_{\rm up}}$ represents the sensing \ac{mi} solely focused on the $k$-th target. Therefore, the upper bound of multi-target sensing \ac{mi}  can be regarded as the sum of individual single-target sensing \ac{mi}.} %when there is no inter-target interference. }
\end{corollary}

As mentioned in \cite{Stoica2007,lijian2007}, the statistical performance of adaptive \ac{mimo} radar heavily relies on the cross-correlation pattern.
It also indicates that the transmit beamforming design should aim to minimize the cross-correlation among signals form specified target directions while optimizing the transmission power in those directions.

Following this principle, in the next section, we use the  upper bound of sensing \ac{mi} as the sensing performance metric to optimize the beamforming vectors, which can potentially enhance transmit power at given target directions\footnote{ While $I_{up}$ depends on the target directions $\theta_k,\forall k$ as can be seen in (\ref{upb_const}), maximizing $I_{up}$ can be interpreted as  optimizing the beamforming vectors $\mathbf{w}$ with respect to the directions towards the potential targets. This is typical for \ac{mimo} radar systems operating in tracking mode, where the BS has knowledge of the target directions and can obtain them from previous observations\cite{liu2020joint}.
In the case of static or slowly moving targets, it is reasonable to utilize the estimated or predicted directions for beamforming design\cite{liu2021cramer}.}.
Concurrently, we use the cross-correlation constraints (\ref{appro_zf_cons}) to ensure the attainability of the upper bound and minimize the cross-correlation pattern.

\section{Joint Beamforming Based on Multi-Objective Optimization} \label{section_optimization}

In this section, we investigate a general transmit  beamforming method that provides a flexible tradeoff between multi-target sensing and multi-user communication performance. %concurrently optimizes multi-target sensing performance and multi-user communication performance.
 We begin by formulating a \ac{moop} to concurrently optimize the sensing \ac{mi} and the \ac{sinr} of each user in Subsection~\ref{formulation}.
Then, we employ the \ac{sdr} to tackle the rank-one constraint in Subsection~\ref{sdr}.
To efficiently solve the complex \ac{moop}, we propose a max-min utility optimization method to attain the Pareto boundary of the \ac{moop} in Subsection~\ref{max-min}, which also considers fairness among multiple users and targets.

%In this section, we 
\subsection{Problem Formulation} \label{formulation}
The expressions in (\ref{sinr_com}) and (\ref{upb_senmi_1}) reveal that the performance of multi-target sensing and multi-user communication are coupled together through transmit beamforming vectors, which can lead to inherent conflicts and tradeoff. 
To comprehensively investigate the tradeoff and optimize the performance of communication and sensing simultaneously, we introduce a \ac{moop} framework. 

Since the maximum sensing \ac{mi} is  achieved by maximizing $I_{\rm up}$  subject to the constraint (\ref{upb_const}), we formulate the \ac{moop} as follows:
\begin{subequations} \label{moop}
	\begin{align}
		\label{moop1_objective}
		\max \limits_{ \left\{\mathbf{w}_{n}\right\} }
		\ & \left\{ {{I}_{\rm up}},{{\gamma }_{1}},\ldots ,{{\gamma }_{{{C}}}} \right\} \\
		{\rm s.t.}
		\label{moop1_pt_constraint}
		\ & \text{tr}\left( \sum\nolimits_{n=1}^{{{C}}+M}{{{\mathbf{w}}_{n}}\mathbf{w}_{n}^{H}}\right)\le {{P}_{T}}, \\
		\label{moop1_miup_constraint}
		\ & {{\mathbf{a}}^{H}}\left( {{\theta }_{j}} \right)\left( \sum\nolimits_{n=1}^{{{C}}+M}{{{\mathbf{w}}_{n}}\mathbf{w}_{n}^{H}}\right)\mathbf{a}\left( {{\theta }_{i}} \right)=0\ \left(i\ne j \right), \\
     	\label{MI_constraint}
     	\ &  {I}_{\rm up}\geq\Lambda, \\
     	\label{sinr_constraint}
     	\ & {{\gamma }_{i}}\ge {{\Gamma }_{i}}, \forall i \in \mathcal{C},  	
	\end{align}
\end{subequations}
where $P_{T}$ is the total transmit power of the BS,  $\Lambda\geq 0$ and ${{\Gamma}_{i}}\geq 0$ indicate the lowest acceptable level of the sensing \ac{mi} and  \ac{sinr} of the $i$-th user, respectively. 
Constraint (\ref{moop1_pt_constraint}) ensures that the total transmit power remains within a predetermined limit, while constraint (\ref{moop1_miup_constraint}) guarantees that the sensing \ac{mi} enables to achieve its upper bound. 
The \ac{moop} \eqref{moop} is non-convex due to its quadratic terms in both objective functions and constraints, making it difficult to solve directly.
Nevertheless, we show in Subsection~\ref{sdr} that it can be reformulated into a more tractable relaxed problem using \ac{sdr}.
In Subsection~\ref{max-min}, we propose a max-min utility optimization method to obtain a specific Pareto optimal solution for the relaxed \ac{moop}. 
Furthermore, we prove that this Pareto optimal solution for the relaxed \ac{moop} also serves as the Pareto optimizer for the original \ac{moop} \eqref{moop}, i.e., the relaxation is tight.

\vspace{-0.5cm}
\subsection{Problem Transformation via Semidefinite Relaxation}\label{sdr} 
\textcolor{black}{In this subsection, we employ the \ac{sdr} strategy to tackle the non-convex problem \eqref{moop}.
To this end, we first introduce  variables ${{\mathbf{R}}_{n}}={{\mathbf{w}}_{n}}\mathbf{w}_{n}^{H}\succeq \mathbf{0}$ with  ${\rm rank} ( \mathbf{R}_{n} ) = 1$, where $n=1,\dots,C+M$.
Subsequently, by substituting $\{\mathbf{R}_{n}\}$ into (\ref{sinr_com}), (\ref{upb_senmi_1}),  (\ref{moop1_pt_constraint}), and (\ref{moop1_miup_constraint}),  we can linearize the quadratic terms in the  problem \eqref{moop}.
}
Thus, the expressions in (\ref{sinr_com}) and (\ref{upb_senmi_1}) are rewritten as 
\begin{equation} 
	\begin{aligned}
		\label{sinr_sdp_objective}
		 {{\gamma }_{i}}=\frac{\mathbf{h}_{i}^{H}{{\mathbf{R}}_{i}}{{\mathbf{h}}_{i}}}{\sum\nolimits_{n=1,n\ne i}^{C}{\mathbf{h}_{i}^{H}{{\mathbf{R}}_{n}}{{\mathbf{h}}_{i}}}\!+\!\sum\nolimits_{n=C+1}^{C+M}{\mathbf{h}_{i}^{H}{{\mathbf{R}}_{n}}{{\mathbf{h}}_{i}}}\!+\!\sigma _{c}^{2}},\forall i \in \mathcal{C},
	\end{aligned}
\end{equation}
\begin{equation}
    \begin{aligned}
        \label{senmi_sdp_objective}
		{{I}_{\rm up}}=\!\sum\nolimits_{{k}=1}^{K}{\log\! \left( 1+\! \frac{{{N}_{r}}\sigma _{k}^{2}L}{\sigma _{r}^{2}}{{\mathbf{a}}^{H}}\!\!\left( {{\theta }_{k}} \right)\!\left(\! \sum\limits_{n=1}^{C+M}{{\mathbf{R}}_{n}} \right)\!\mathbf{a}\left( {{\theta }_{k}} \right) \right)}.
    \end{aligned}
\end{equation}
The constraints (\ref{moop1_pt_constraint}) and (\ref{moop1_miup_constraint}) are rewritten as
\begin{subequations} 
	\begin{align}
		\label{moop1_pt_constraint_n}
		\ & \text{tr}\left( \sum\nolimits_{n=1}^{{{C}}+M}{{{\mathbf{R}}_{n}}} \right)\le {{P}_{T}}, \\
		\label{moop1_miup_constraint_n}
		\ & \text{tr}\left( \mathbf{a}\left( {{\theta }_{i}} \right){{\mathbf{a}}^{H}}\!\left( {{\theta }_{j}} \right)\!\left( \sum\nolimits_{n=1}^{{{C}}+M}\!{{{\mathbf{R}}_{n}}} \right) \right)=0\ \left(i\ne j \right).
	\end{align}
\end{subequations}

With the newly derived objective functions and constraints, the \ac{moop}  (\ref{moop}) is  reformulated as
\begin{subequations} \label{moop2}
	\begin{align}
		\label{moop2_objective}
		\max \limits_{ \left\{\mathbf{R}_{n}\right\} }
		\ & \left\{ {{I}_{\rm up}},{{\gamma }_{1}},\ldots ,{{\gamma }_{{{C}}}} \right\} \\
		{\rm s.t.}
		\label{moop2_pt_constraint}
		\ & \text{tr}\left( \sum\nolimits_{n=1}^{{{C}}+M}{{{\mathbf{R}}_{n}}} \right)\le {{P}_{T}}, \\
		\label{moop2_miup_constraint}
		\ & \text{tr}\left(\! \mathbf{a}\left( {{\theta }_{i}} \right){{\mathbf{a}}^{H}}\!\left( {{\theta }_{j}} \right)\!\left( \sum\nolimits_{n=1}^{{{C}}+M}\!{{{\mathbf{R}}_{n}}}\!  \right) \! \right)=0\ \! \left(i\ne j \right),\\
		\label{semi_constraint}
		\ & {\mathbf{R}}_{n}\succeq 0, \ \forall n,\\
		\label{moop2_rank1_constraint}
		\ &  {\rm rank} ( {\mathbf{R}}_{n} ) = 1,\ \forall n,\\
		\label{moop2_misinr_constraints1}
		\ & {\rm (\ref{MI_constraint})\ and \ (\ref{sinr_constraint})}.
	\end{align}
\end{subequations}

Since the rank-one constraints (\ref{moop2_rank1_constraint}) are non-convex in problem (\ref{moop2}), we use the \ac{sdr} strategy to tackle it.  Omitting the rank-one constraints leads to the following relaxation:
\begin{subequations} \label{moop3}
	\begin{align}
		\label{moop2_objective2}
		\max \limits_{ \left\{\mathbf{R}_{n}\right\} }
		\ & \left\{ {{I}_{\rm up}},{{\gamma }_{1}},\ldots ,{{\gamma }_{{{C}}}} \right\} \\
		{\rm s.t.}
	%	\label{moop2_semi_constraint}
	%	\ & {\mathbf{R}}_{n}\succeq 0, \ \forall n, \\
		\label{moop2_misinr_constraints3}
		\ & {\rm (\ref{MI_constraint})\ , (\ref{sinr_constraint})\ , (\ref{moop2_pt_constraint})\ , (\ref{moop2_miup_constraint})\ and \ (\ref{semi_constraint})}.
	\end{align}
\end{subequations}
The \ac{moop} (\ref{moop3}) aims to find a transmit strategy $\left\{\mathbf{R}_{n}\right\}$ that satisfies transmit power and cross-correlation constraints while maximizing sensing performance $I_{\text{up}}$ and communication performance $\gamma_{1},\cdots,\gamma_{C}$ for all users.
Nevertheless, the conflict arises between maximizing $\gamma_{1},\cdots,\gamma_{C}$ and $I_{\text{up}}$, as they are coupled through the transmit strategy $\left\{\mathbf{R}_{n}\right\}$.
Typically, there does not exist a singular transmit strategy that can simultaneously optimize all these objectives.
Therefore, it is instructive to consider the set of feasible performance outcomes for all feasible transmit strategies, i.e., the achievable performance region of the multi-target and multi-user \ac{isac} system.
The achievable performance region $\mathcal{M}\subset \mathbb{R}_{+}^{{{C}}+1}$ is defined as a set of achievable performance pairs with all the feasible transmit strategies: %[TSP-2013] Pareto boundary
\begin{equation} 
	\label{apr_M}
	\begin{aligned}		
		\mathcal{M}=\left\{ \left( {{I}_{\rm up}},{{\gamma }_{1}},\cdots ,{{\gamma }_{{{C}}}} \right):\forall \left\{ {{\mathbf{R}}_{1}},\ldots ,{{\mathbf{R}}_{{{C}}+M}} \right\}\in \mathcal{R} \right\},
	\end{aligned}
\end{equation}
where $\mathcal{R}$ is the feasible transmit strategy set with transmit power and zero-forced cross-correlation constraints:
\begin{equation}
	\label{fsr}
	\begin{aligned}
		\mathcal{R}=& \!\left\{\!\left( {{\mathbf{R}}_{1}},\ldots ,{{\mathbf{R}}_{{{C}}+M}} \right)\!:\!{{\mathbf{R}}_{n}}\succeq 0,\!\ \forall n,\text{tr}\left(\!\sum\nolimits_{n=1}^{{{C}}+M}\!\!{{{\mathbf{R}}_{n}}}\! \right)\!\le\! {{P}_{T}}, \right. \\ 
		& \left. \text{tr}\left( \mathbf{a}\left( {{\theta }_{i}} \right){{\mathbf{a}}^{H}}\left( {{\theta }_{j}} \right)\left( \sum\nolimits_{n=1}^{{{C}}+M}{{{\mathbf{R}}_{n}}} \right) \right)=0\text{ }\left(i\ne j \right) \right\}.		
	\end{aligned}
\end{equation}
The region describes the achievable sensing \ac{mi} and  communication \ac{sinr}s that can be simultaneously attained under transmit power and the zero-forced cross-correlation constraints.
Typically, we seek to obtain a set of optimal solutions for \ac{moop},  known as Pareto optimal solutions, which are incomparable to each other and  no superior solution exists in the objective space. 
These solutions are found on a subset of the outer boundary of $\mathcal{M}$ referred to as the Pareto boundary, where an improvement in a single performance implies a degradation in the other performances:
 \begin{definition} \label{definition_Pareto_boundary}
	\emph{ The \emph{Pareto boundary} $\partial {{\mathcal{M}}}\subseteq \mathcal{M}$ consists of all $\mathbf{x}\in \mathcal{M}$ for which there is no $\mathbf{x}^{\prime}\in \mathcal{M}\backslash \left\{ \mathbf{x} \right\}$ with  $\mathbf{x}^{\prime}\geq \mathbf{x}$\cite[Definition 4]{bjornson2012Pareto}}.
\end{definition}

\subsection{Max-Min Utility Optimization}\label{max-min}
We next provide the algorithm to obtain the Pareto optimal solutions for the \ac{moop} (\ref{moop3}).
A common approach to  address the \ac{moop} is to combine multiple objective functions into a system utility function, denoted by  $H(\cdot)$.
In order to meet the minimum communication and sensing performance thresholds and  account for fairness between the multi-target sensing performance and multi-user communication performance, we employ the  minimum weighted compromise function as the system utility function\cite{bjornson2012robust,bjornson2012Pareto}.
For a given feasible operating point $\mathbf{x}=({{I}_{\rm up}},{{\gamma }_{1}},\cdots ,{{\gamma }_{{{C}}}})\in \mathcal{M}$, the system utility function is given by
\begin{equation}
		\label{max-min-fair}
	H\left( \mathbf{x} \right)=\underset{i}{\mathop{\min }}\,\left\{ \frac{{{I}_{\rm up}}-{\Lambda}}{\alpha },\frac{{{\gamma }_{i}}-{{ \Gamma}_{i}}}{{{\omega }_{i}}} \right\},
	\end{equation}
where the communication and sensing performance should be in excess of the lowest acceptable levels, i.e., ${I}_{\rm up}\geq\Lambda$ and ${\gamma }_{i}\geq{\Gamma}_{i}, \forall i \in \mathcal{C}$.
The parameters $\alpha \in {{\mathbb{R}}_{+}}$ and ${{\omega }_{i}}\in {{\mathbb{R}}_{+}}$ are the weights of sensing and $i$-th communication user, respectively, which satisfy $\sum\nolimits_{i=1}^{{{C}}}{{{\omega }_{i}}}+\alpha =1$.
Next, we demonstrate that a set of Pareto solutions for the \ac{moop} (\ref{moop3}) can be obtained by solving the following system function optimization problem:
\begin{subequations} \label{moop4}
	\begin{align}
		\max \limits_{ \left\{\mathbf{R}_{n}\right\} }
		\ & \min \limits_{i} 
		\ \left\{ \frac{{{I}_{\rm up}}-\Lambda}{\alpha },\frac{{{\gamma }_{i}}-{{ \Gamma}_{i}}}{{{\omega }_{i}}} \right\} \\
		{\rm s.t.}
		\label{moop4_sub_1_constraints1}
		\ & {\rm (\ref{MI_constraint})\ ,(\ref{sinr_constraint})\  ,(\ref{moop2_pt_constraint}) \ ,(\ref{moop2_miup_constraint})\ and \ (\ref{semi_constraint})}.
	\end{align}
\end{subequations}
It is worth noting that the objective function in (\ref{moop4}) can provide  a flexible tradeoff between sensing and communication by assigning appropriate weights.
For instance, in two extreme weight configuration cases, problem (\ref{moop4}) can be transformed  into the sensing \ac{mi} optimization problem with  \ac{sinr} constraints (denoted by Case 1) and the  \ac{sinr} optimization problem with the sensing \ac{mi} constraint (denoted by Case 2), as follows: 
\begin{case} \label{case1}
	\emph{ In this case, we set $\Lambda =0$, $\alpha =1$, and ${{\omega }_{i}}\to 0$, which gives higher priority to sensing over communication.  
    The problem (\ref{moop4}) is transformed into an optimization problem that maximizes the sensing \ac{mi} while satisfying each user's \ac{sinr} constraint. % refer to[TSP]chenli
		\begin{subequations} \label{moop_sub_1}
			\begin{align}
				\label{moop_sub_1_objective}
				\max \limits_{ \left\{\mathbf{R}_{n}\right\} } 
				\ & \left\{ {I}_{\rm up} \right\} \\
				{\rm s.t.}
				\label{moop_sub_1_constraints1}
				\ & {\rm (\ref{sinr_constraint})\ ,(\ref{moop2_pt_constraint}) \ ,(\ref{moop2_miup_constraint})\ and \ (\ref{semi_constraint})}.
			\end{align}
		\end{subequations}
	}
\end{case}

\begin{case} \label{case2}
	\emph{ In this case, we set ${\Gamma}_{i} =0$, ${{\omega}_{i}}=1/C$, and $\alpha\to 0$,  which gives higher priority to communication. 
   Additionally, every user is considered to have the same importance. 
   As a result, problem (\ref{moop4}) reduces to a maximization of the minimum \ac{sinr} subject to the sensing \ac{mi} constraint. % refer to[TSP]chenli
		\begin{subequations} \label{moop_sub_2}
			\begin{align}
				\label{moop_sub_2_objective}
				\max \limits_{ \left\{\mathbf{R}_{n}\right\} } \min \limits_{i} 
				\ & \left\{ {{\gamma }_{i}} \right\} \\
				{\rm s.t.}
				%\label{moop_sub_2_constraints2}
				%\ & {{I}_{\rm up}}\ge \Lambda ,\\
				\label{moop_sub_2_constraints1}
				\ &{\rm (\ref{MI_constraint})\ ,(\ref{moop2_pt_constraint}) \ ,(\ref{moop2_miup_constraint})\ and \ (\ref{semi_constraint})}.
			\end{align}
		\end{subequations}
	}
\end{case}

However, the achievable performance region $\mathcal{M}$ is  typically non-convex, which poses a challenge in obtaining Pareto solutions for the \ac{moop} (\ref{moop3}).
For instance, if $\mathcal{M}$ is a non-normal set with internal holes, it can lead to a complex and challenging resolution of the Pareto boundary\cite{bjornson2012Pareto}.
Given this, we first investigate the properties of $\mathcal{M}$ to demonstrate its compactness and normality, and consequently demonstrate that the Pareto solutions for problem (\ref{moop3}) can be attained by solving the system function optimization problem (\ref{moop4}).
\begin{definition} \label{definition_compact_set}
	\emph{ $\mathcal{M}\subset \mathbb{R}_{+}^{n}$ is called a compact set if it is closed and bounded\cite[Def. 2]{chen2021joint}.}
\end{definition}
\begin{definition} \label{definition_normal_set}
	\emph{ $\mathcal{M}\subset \mathbb{R}_{+}^{n}$ is called a normal set if for any point $\mathbf{x}\in \mathcal{M}$, all $\mathbf{x}^{\prime}\in \mathbb{R}_{+}^{n}$ with $\mathbf{x}^{\prime}\le \mathbf{x}$ (component-wise inequalities) also satisfy $\mathbf{x}^{\prime}\in \mathcal{M}$\cite[Def. 4]{bjornson2012robust}.}
\end{definition}

\begin{lemma} \label{lemma_1}
	\emph{The performance region $\mathcal{M}$ is a compact and normal set.}
\end{lemma}
\begin{IEEEproof}
	Please refer to Appendix \ref{appc}.
\end{IEEEproof}

Additionally,  we can observe from  (\ref{max-min-fair}) that the system utility function $H(\cdot)$ is strictly increasing.
Combining this with the characterization of  $\mathcal{M}$  as a compact and normal set, we can derive the following important conclusion.
\begin{lemma} \label{lemma_pro_equ}
	\emph{ If $H(\cdot)$ is a strictly increasing function and $\mathcal{M}$ is a compact and normal set, the global optimum to $\max \limits_{ \mathbf{x}\in\mathcal{M} }
		\  H(\mathbf{x})$  is attained on $\partial {{\mathcal{M}}}$. }
\end{lemma}
\begin{IEEEproof}
	Please refer to \cite[Lem. 2]{bjornson2012robust}.
\end{IEEEproof}
Based on Lemma \ref{lemma_pro_equ}, we can attain the Pareto boundary of \ac{moop} (\ref{moop3}) by solving the problem  (\ref{moop4}).

We next present the algorithm to solve the problem (\ref{moop4}). Specifically, letting
\begin{equation} \label{r}
	r =\underset{i}{\mathop{\min }}\,\left\{ \frac{{{I}_{\rm up}}-\Lambda}{\alpha },\frac{{{\gamma }_{i}}-{{ \Gamma}_{i}}}{{{\omega }_{i}}} \right\},
\end{equation}
the max-min utility problem (\ref{moop4}) can be recast as 
\begin{subequations} \label{moop_4}
	\begin{align}
		\label{moop4_objective}
		\max \limits_{ \left\{\mathbf{R}_{n}\right\} } 
		\ & \ \ r\\
		{\rm s.t.}
		\label{moop4_constraints2}
		\ & {{I}_{\rm up}}\ge \Lambda+\alpha r,\\
		\label{moop4_constraints3}
		\ & {{\gamma }_{i}}\ge {{\Gamma }_{i}}+\omega_{i} r, \forall i \in \mathcal{C},\\
		\label{moop4_constraints1}
		\ & {\rm (\ref{moop2_pt_constraint}) \ ,(\ref{moop2_miup_constraint})\ and \ (\ref{semi_constraint})}.
	\end{align}
\end{subequations}

We observe that the
 problem (\ref{moop_4}) has a geometric interpretation, where we can find the optimal point in the Pareto boundary $\partial {{\mathcal{M}}}$ by starting from an initial point $\mathbf{x}=\left(\Lambda,\Gamma_{1},\cdots,\Gamma_{C},\right)\in \mathcal{M}$ determined by the performance thresholds and following a ray in the direction of $\mathbf{c}=\left[ \alpha ,{{\omega }_{1}},\ldots ,{{\omega }_{{{C}}}} \right]^{T}$ determined by the sensing and communication weights.
Since $\mathcal{M}$ is a compact and normal set, the ray intersects the Pareto boundary at a unique point.
For fixed weights $ \alpha ,{{\omega }_{1}},\ldots, {{\omega }_{{{C}}}} $, we first define an upper bound of $r$ denoted by $r_{\rm max}$, where $\mathbf{x}+\mathbf{c}r_{\rm max}$ is outside $\mathcal{M}$. 
The initial upper bound for $r$ can be chosen as a sufficiently large number, or it can be computed as $r_{\rm max}=\sum_{i=1}^C \frac{P_T {\omega }_{i}\left\|h_i\right\|^2}{\sigma_i^2}+\alpha \sum_{k=1}^K \log \left(1+\frac{P_T N_t^2 \sigma_k^2 L}{\sigma_r^2}\right)$.
The optimal value of the problem (\ref{moop_4}) lies on the line segment $\left[0,{{r}_{\max }}\right]$.

Therefore, the  problem (\ref{moop_4}) can be solved  by performing bisection search on the range $\left[0,{{r}_{\max }}\right]$, which decomposes it into a series of feasible subproblems.
That is, for a given $r\ge 0$, we can efficiently check if there exists the feasible transmit strategies $\left\{ {\mathbf{R}}_{1},\ldots ,{\mathbf{R}}_{C+M} \right\}$ that satisfies the constraints in (\ref{moop_4}).
However, due to the non-convex and nonlinear constraint (\ref{moop4_constraints2}), the feasibility problem remains computationally intractable.
To tackle this issue, we further consider fairness between multiple targets.
By assigning weights ${\xi}_k$ to each sensing target, where $\sum\nolimits_{k=1}^{K}{\xi}_k =1$, we can equivalently transform the non-convex constraint (\ref{moop4_constraints2})  into a series of convex constraints. 
Therefore, we can achieve fairness in the sensing performance of each  target and convert the feasibility problem into the following convex problem:
%Thus, the convex feasibility problem can be formulated as
\begin{subequations} \label{find_feasible}
	\begin{align}
		\label{f_f_objective}
		 \text{find}
		 \ & \left\{ {{\mathbf{R}}_{1}},\ldots ,{{\mathbf{R}}_{{{C}}+M}}\right\} \\ 
		{\rm s.t.}
		\label{f_f_tk}
		\ & \log \left( 1+{\text{SINR}_{k}} \right) \ge {\xi}_k\left(\Gamma+\alpha r\right),\forall k, \\
		\label{f_f_tksum}
		%\ & \sum\nolimits_{k=1}^{K}{{{t}_{k}}}\ge \Lambda+\alpha r,\\
		%\label{ff_constraints1}
		\ & {\rm (\ref{moop2_pt_constraint}) \ ,(\ref{moop2_miup_constraint})\  ,(\ref{semi_constraint})\ and \
			(\ref{moop4_constraints3})	}.
	\end{align}
\end{subequations}

We summarize the procedure to solve the problem (\ref{moop_4}) in Algorithm \ref{algorithm_SDR_bisection}.
In particular, for a given convergence threshold $\varepsilon$, the algorithm can find an interval $\left[{{r}_{\min }},{{r}_{\max }}\right]$ for the optimal value of (\ref{moop_4}) that satisfies $\left| {{r}_{\min }},{{r}_{\max }}\right|\le \varepsilon$ in a limited number of iterations, which scales only with $\varepsilon$, i.e., $\left\lceil\log_2\left({{r}_{\max }}/\varepsilon\right)\right\rceil$.
\begin{algorithm}[t]
	\caption{Bisection Search to Attain the Pareto Boundary}
	\label{algorithm_SDR_bisection}
	\begin{algorithmic}[1]
		\renewcommand{\algorithmicrequire}{\textbf{Initialize}}
		\renewcommand{\algorithmicensure}{\textbf{Output}}
		\STATE \textbf{Initialize} $\Lambda$, $\alpha$, $\{{\Gamma}_{i}\}$, $\{{\omega }_{i}\}$, ${{r}_{\min }}$, ${{r}_{\max }}$,  %where the optimal $r_{opt}$ lies in $[{{r}_{\min }},{{r}_{\max }}]$,  
		${{r}^{(1)}}=({{{r}_{\min }}+{{r}_{\max }}})/{2}$, the tolerance $\varepsilon$,  and set $j=1$.
		\REPEAT \label{alg1_start}
		\IF  {the problem (\ref{find_feasible}) is feasible for ${{r}^{(j)}}$ }
		\STATE ${{r}_{\min }}={{r}^{(j) }}$;
		\ELSE
		\STATE ${{r}_{\max }}={{r}^{(j)}}$;
		\ENDIF
		\STATE Update $j := j+1$;
		\STATE Update ${{r}^{(j) }}:= [{{r}}_{\min }+{{r}_{\max }}]/{2}$;
		\UNTIL $\left| {{r}^{(j)}}-{{r}^{(j-1) }} \right|\le \varepsilon$;\label{alg1_end}
		%\STATE Fix the performance constraints according to $\mathbf{g}\left( {{d}^{\prime }} \right)$;
		\STATE Solve the feasible problem (\ref{find_feasible}) for ${r}^{(j)}$. 
		\STATE \textbf{Output} the optimal solution $\left\{ {{\mathbf{R}}_{1}},\ldots ,{{\mathbf{R}}_{{{C}}+M}} \right\}$.
	\end{algorithmic}
\end{algorithm}
Specifically, when the convergence threshold $\varepsilon$ is small enough, it is reasonable to approximate that the solution obtained by Algorithm \ref{algorithm_SDR_bisection} is the optimal solution of the problem (\ref{moop_4}).
\textcolor{black}{The bisection search method in Algorithm \ref{algorithm_SDR_bisection} requires $\left\lceil\log_2\left({r_{\max}}/{\varepsilon}\right)\right\rceil$ iterations to find an interval of length $\varepsilon$ containing the optimal value.
In each iteration, the convex feasibility problem \eqref{find_feasible} is solved using interior-point methods, which has a worst-case complexity of $\mathcal{O}\left((C+M)^3N_t^6+(K^2+C)(C+M)N_t^2\right)$\cite{ye2011interior}, where $C$, $M$, $K$, and $N_t$ denote the number of communication users, radar streams, targets, and transmit antennas, respectively.
Therefore, the overall computational complexity of Algorithm 1 is given by $\mathcal{O}\left(\left\lceil\log_2\left(\frac{r_{\max}}{\varepsilon}\right)\right\rceil\cdot\left((C+M)^3N_t^6+(K^2+C)(C+M)N_t^2\right)\right)$, which is polynomial in the number of iterations, targets, users, radar streams, and antennas.}

Based on the previous analysis, it can be concluded that the optimal solution to problem (\ref{moop_4}) lies on the Pareto boundary of the problem (\ref{moop3}).
However, the  optimal solution of the problem (\ref{moop_4}) may be with high ranks, indicating that the \ac{sdr} solution is not necessarily tight to (\ref{moop2}).
We introduce Theorem~\ref{lemma_2} to prove the existence of an optimal rank-one solution for problem (\ref{moop_4}),   which corresponds to the rank-one Pareto optimal solution for problem (\ref{moop3}).
\begin{theorem} \label{lemma_2}
	\emph{The problem (\ref{moop_4}) always has an optimal solution $\left\{ {{{\mathbf{\bar{R}}}}_{1}},\ldots, {{{\mathbf{\bar{R}}}}_{{C}+M}} \right\}$ that satisfies
		\begin{equation}
			\label{rank1_opt}
			\begin{aligned}
				{\rm rank} ( {{{\mathbf{\bar{R}}}}_{n}} ) = 1\ , \forall n .	  		 
			\end{aligned}
		\end{equation}}	
\end{theorem}

\begin{IEEEproof}
	Please refer to Appendix \ref{appd}.
\end{IEEEproof}

We next introduce the construction process of the rank-one solution. 
According to Appendix~\ref{appd}, when the optimal solution for problem (\ref{moop_4}) is obtained as $\left\{ {{\mathbf{R}}_{1}^{*}},\ldots, {\mathbf{R}}_{C+M}^{*} \right\}$, we can use it to construct a rank-one optimal solution $\left\{ {{{\mathbf{\bar{R}}}}_{1}},\ldots, {{{\mathbf{\bar{R}}}}_{{C}+M}} \right\}$ and the corresponding optimal beamforming vectors $\left\{ {{{\mathbf{\bar{w}}}}_{1}},\ldots, {{{\mathbf{\bar{w}}}}_{{C}+M}} \right\}$.
Firstly, $\left\{ {{{\mathbf{\bar{R}}}}_{1}},\ldots, {{{\mathbf{\bar{R}}}}_{{C}}} \right\}$ and $\left\{ {{{\mathbf{\bar{w}}}}_{1}},\ldots, {{{\mathbf{\bar{w}}}}_{{C}}} \right\}$ can be computed as
\begin{equation}
	\label{rankone-vec1}
	\begin{aligned}			{{\mathbf{\bar{w}}}_{i}}=\frac{\mathbf{R}_{i}^{*}{{\mathbf{h}}_{i}}}{\sqrt{\mathbf{h}_{i}^{H}\mathbf{R}_{i}^{*}{{\mathbf{h}}_{i}}}},  {{\mathbf{\bar{R}}}_{i}}={{\mathbf{\bar{w}}}_{i}}\mathbf{\bar{w}}_{i}^{H}, \forall i \in \mathcal{C}.
	\end{aligned}
\end{equation}
Then, we can obtain $\left\{ {{{\mathbf{\bar{w}}}}_{C+1}},\ldots, {{{\mathbf{\bar{w}}}}_{{C+M}}} \right\}$ by taking
Cholesky decomposition:
\begin{equation}
	\label{rankone-vec2}
	\begin{aligned}		
{{\mathbf{\bar{W}}}_{r}}\mathbf{\bar{W}}_{r}^{H}=\sum\nolimits_{n=1}^{C+M}{\mathbf{R}_{n}^{*}}-\sum\nolimits_{i=1}^{C}{{{{\mathbf{\bar{R}}}}_{i}}},
	\end{aligned}
\end{equation}
where ${{\mathbf{\bar{W}}}_{r}}=\left[ {{{\mathbf{\bar{w}}}}_{{C}+1}},\ldots ,{{{\mathbf{\bar{w}}}}_{{C}+M}} \right]$ is a lower triangular matrix.
Therefore, the rank-one matrices  $\left\{ {{{\mathbf{\bar{R}}}}_{C+1}},\ldots, {{{\mathbf{\bar{R}}}}_{{C+M}}} \right\}$ can be constructed as ${{\mathbf{\bar{R}}}_{j}}={{\mathbf{\bar{w}}}_{j}}\mathbf{\bar{w}}_{j}^{H}$ for $j=C+1,\ldots ,C+M$.
 Hence, $\left\{ {{{\mathbf{\bar{R}}}}_{1}},\ldots, {{{\mathbf{\bar{R}}}}_{{C}+M}} \right\}$ is a Pareto optimal solution to \eqref{moop2},  which demonstrates that it is tight to  transform (\ref{moop2}) into  (\ref{moop3}) via  \ac{sdr}.
Furthermore, the beamforming vectors $\left\{{{{\mathbf{\bar{w}}}}_{1}}, \ldots, {{{\mathbf{\bar{w}}}}_{{C}+M}} \right\}$ constitute a Pareto optimal solution for the original \ac{moop} \eqref{moop}.

\section{Simulation Results} \label{section_simulation}
In this section, we evaluate the performance of the proposed multi-objective optimization for the \ac{mimo}-\ac{isac} systems through the Monte-Carlo simulation results.
The communication performance is evaluated in terms of the average rate of the \ac{mu-mimo} communication defined in (\ref{averate_com}), while the sensing performance is evaluated in terms of the sensing \ac{mi}. 
We use the following simulation settings unless specified otherwise.
The BS is equipped with $N_{t}=32$ transmit antennas and $N_{r}=32$ receive antennas.
The length of the transmit signal is set to $L=1024$\cite{liu2020joint}.
Both antenna arrays are ULAs with the same antenna spacing $d = \lambda/2$.
The total transmit power is $P_{T}=40{\rm \ dBm}$.
We assume that the noise power for each communication user are the same, i.e., ${\sigma _{i}^{2}}=0{\rm \ dBm}, \forall i \in \mathcal{C}$\cite{liu2021cramer}.
And the communication \ac{snr} of each user is defined as ${\text{SNR}}=\frac{ P_T}{\sigma_i^2}$.
We assume that the noise power in the received radar signal is ${\sigma _{r}^{2}}=0{\rm \ dBm}$\cite{liu2021cramer}. For simplicity, the variance of the scattering coefficients are assumed to be the same, i.e., $\sigma _{k}^{2}=\sigma^{2}, \forall k $. 
The radar \ac{snr} is defined as $ {\text{SNR}_\text{radar}}=\frac{|\sigma|^2 P_T}{\sigma_r^2}$.
Without loss of generality, we consider Rician fading for the communication channels.
In this case, the communication channel for each user has the structure\cite{sen2022frequency}
\begin{equation}
\begin{aligned}
{{\mathbf{h}}_{i}}=\sqrt{\frac{\mu }{\mu +1}}{\Delta }+\sqrt{\frac{1}{\mu +1}}\mathbf{u}	,\forall i \in \mathcal{C},	
\end{aligned}
\end{equation}
where ${\boldsymbol{\Delta }}$ denotes the complex line-of-sight phase vector with the $n$-th element having the property $\left|{\Delta} _n\right|^2=1$, $\mathbf{u}$ denotes the scattered component vector with the $n$-th element $u_n \sim \mathcal{C N}(0,1)$, and $\mu$ is the Rician factor.
Especially  when $\mu= 0$, the Rician fading channel degenerates to the Rayleigh fading channel\cite{wen2011sum}.
The convergence tolerance is set to $\epsilon = 0.01$.

To validate the effectiveness of our proposed beamforming design based on multi-objective optimization (labeled `Proposed method'), the following schemes are considered as benchmarks:
\begin{itemize}
	\item[\rmnum{1}.] \textbf{Radar-only}: It denotes the scheme without considering communication  \ac{sinr} constraints, which helps evaluate the best sensing performance and serves as the performance upper bound of radar sensing.
	\item[\rmnum{2}.] \textbf{Communication-only}: This scheme only considers the data transmission by omitting the sensing \ac{mi} constraints 
(\ref{moop4_constraints2}) when solving the problem  (\ref{moop_4}). This scheme can yield maximum communication performance.   
	\item[\rmnum{3}.] \textbf{\ac{mi}-constrained}: This scheme only exploits the downlink communication signals for sensing, and maximizes the lower bound of communications \ac{sinr} under the sensing \ac{mi} constraint\cite{ni2021multi}. 
        \item[\rmnum{4}.] \textbf{Sensing-centric}: This scheme corresponds to Case~\ref{case1}, which maximizes the sensing \ac{mi} while satisfying the \ac{sinr} constraints for each user.
         \item[\rmnum{5}.] \textbf{Communication-centric}: This scheme corresponds to Case~\ref{case2}, which maximizes the \ac{sinr} for each user while satisfying the constraints of sensing \ac{mi}.
         \item[\rmnum{6}.] \textbf{ZF-violated}: This scheme maximizes the upper bound of sensing \ac{mi} and the \ac{sinr} for each user while neglecting the zero-forced cross-correlation constraints in (\ref{moop2_miup_constraint}).

\end{itemize}

\subsection{Convergence Performance}
\begin{figure} [t!]
	\centering
	\subfloat[The sensing \ac{mi} versus the number of iterations for $N_{t}=8, 16, 24, 32$.]{\label{convergence MI}
		\includegraphics[width=0.35 \textwidth]{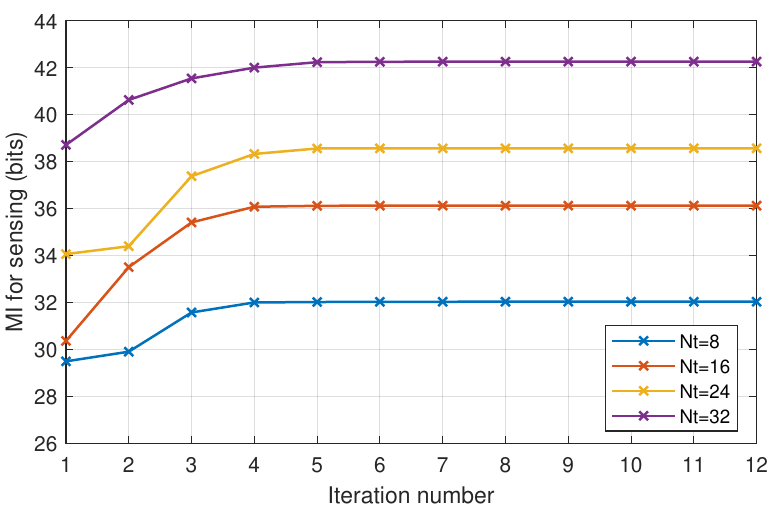}}\\
	\subfloat[The average rate of communication versus the number of iterations for $N_{t}=8, 16, 24, 32$.]{\label{convergence SINR}
		\includegraphics[width=0.35 \textwidth]{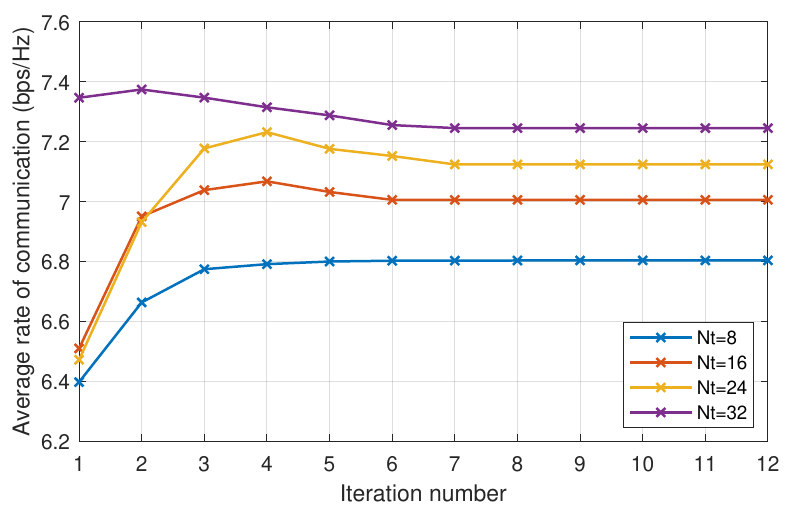} }
	\caption{Convergence of the proposed algorithm. }
	\label{convergence} 
\end{figure}
This subsection aims at analyzing the convergence performance of the proposed algorithm considering different numbers of antennas.
We assume that the number of receive antennas $N_{r}$ is set equal to $N_{t}$.
The numbers of communication users and sensing targets are $C=2$ and $K=2$, respectively. 

Fig. \ref{convergence}\subref{convergence MI} shows the sensing \ac{mi} versus iteration number.
It can be observed that the sensing \ac{mi} increases along with the iteration monotonously for all considered conditions.  
It is noted that the average number of iterations that makes the sensing \ac{mi} converge is about 5.
The average rate of communication versus iteration number is also depicted in Fig.~\ref{convergence}\subref{convergence SINR}.
Besides, it can be observed in Fig. \ref{convergence}\subref{convergence MI}  and Fig. \ref{convergence}\subref{convergence SINR} that  the number of iterations required in the algorithm increases with the increase of $N_t$, because the increase of antenna numbers enlarges the feasible set  $\left\{ {\mathbf{R}}_{1},\ldots ,{\mathbf{R}}_{C+M} \right\}$.

\subsection{Beampattern Performance}
\begin{figure} [t]
	\centering
	\subfloat[The optimized transmit beampatterns with $K=3$.]{\label{beampattern_k3} 
		\includegraphics[width=0.45 \textwidth]{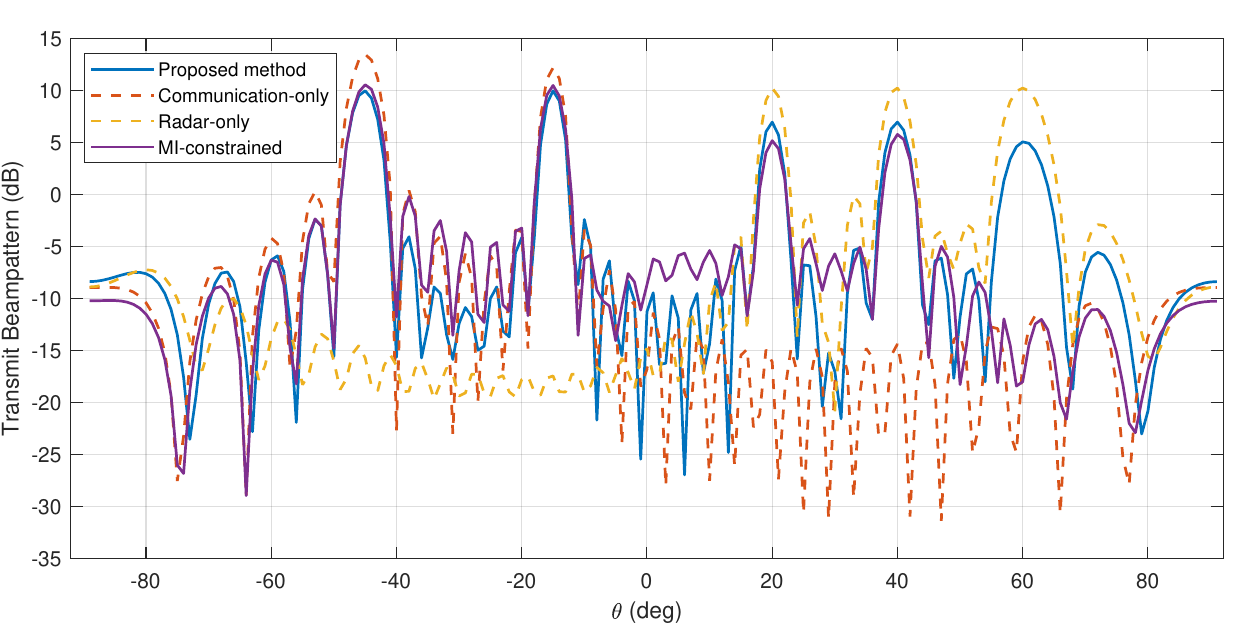}}\\
	\subfloat[The optimized transmit beampatterns with $K=2$.]{\label{beampattern_k2}
		\includegraphics[width=0.45 \textwidth]{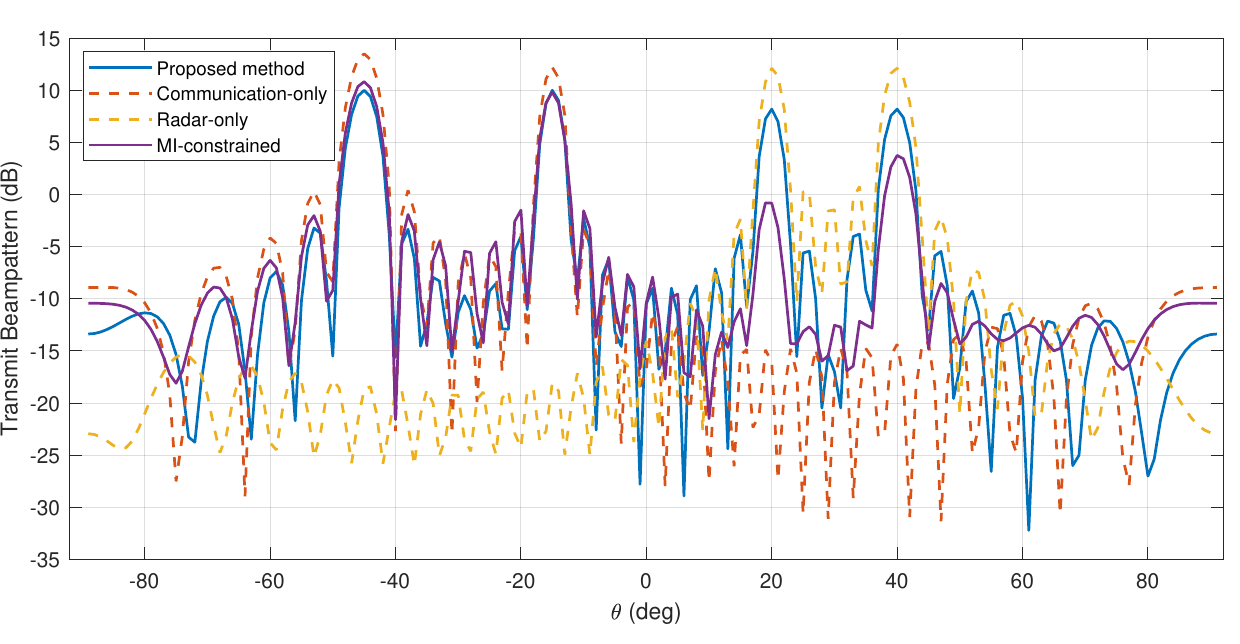} }
	\caption{The transmit beampattern for the proposed method and benchmarks. }
	\label{beampattern} 
\end{figure}

In this subsection, we show the optimized beampattern performance for different $K$ values.
We assume $C=2$ users at the location $\theta_{C1}=-45^{\circ}$ and $\theta_{C2}=-15^{\circ}$, respectively.
The  \ac{sinr} constraint of each communication user is ${{\Gamma }_{i}}=5 {\rm \ dB}, \forall i \in \mathcal{C}$.
And the threshold of  sensing \ac{mi} is $\Lambda = 10{\rm \ bits}$.
The beampatterns obtained via the mentioned schemes are depicted in Fig. \ref{beampattern}\subref{beampattern_k3} for $K=3$ targets located at $20^{\circ}$, $40^{\circ}$, $60^{\circ}$ and in Fig. \ref{beampattern}\subref{beampattern_k2} for $K=2$ targets located at $20^{\circ}$, $40^{\circ}$.

It can be seen from Fig. \ref{beampattern}\subref{beampattern_k3} that the proposed method has three radar mainlobes, matching  the locations of the Radar-only design.
However, there are only two radar mainlobes located at around $20^{\circ}$ and  $40^{\circ}$ for the  \ac{mi}-constrained approach.
We note that the \ac{dof} for \ac{mimo} beamforming is determined by the rank of the covariance matrix\cite{liu2020joint}.  
In the  \ac{mi}-constrained approach, only $C=2$ communication streams are transmitted, and thus the \ac{dof} is no larger than the number of users $C$.
If $C$ is less than the rank of the optimal Radar-only design covariance matrix,  the   \ac{mi}-constrained approach cannot provide enough \ac{dof} to synthesize the desired radar beams, explaining the degraded beampattern observed in Fig.~\ref{beampattern}\subref{beampattern_k3}.
The proposed method simultaneously transmits $C$ transmission streams and $M$ probing streams, which can provide enough  \ac{dof} for transmit beamforming, and thus can produce the beampatterns close to that of the optimal Radar-only method.
In Fig. \ref{beampattern}\subref{beampattern_k2}, however, it is clearly observed that the proposed approach  has enough \ac{dof} to generate a beampattern similar to that of the  Radar-only method for $K=2$ targets. 

%Observing the results in Fig. \ref{beampattern}\subref{beampattern_k3} and Fig. \ref{beampattern}\subref{beampattern_k2}, the power of the mainlobe generated by the proposed method is better than the  \ac{mi}-constrained approach.
%

\subsection{Comparisons of Communication and Sensing Performance}

\begin{figure} [t]
	\centering	\includegraphics[width=0.4\textwidth]{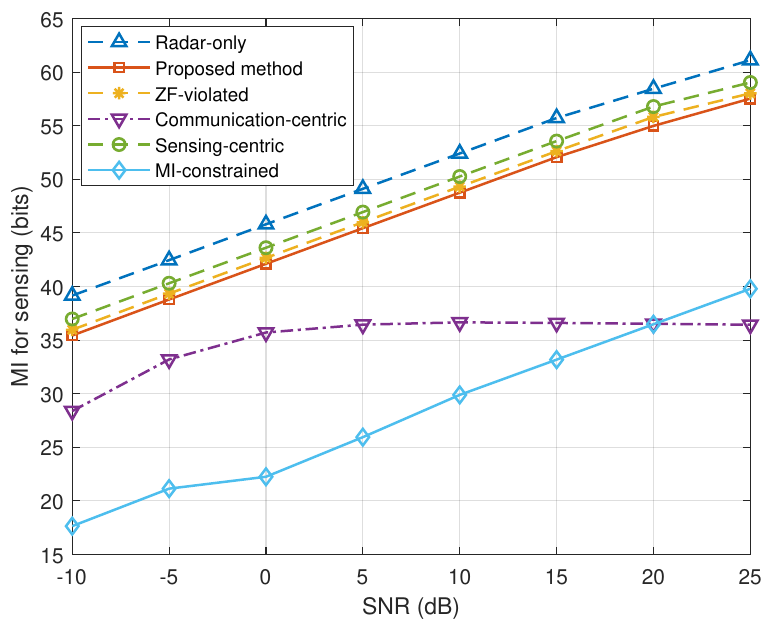}
	\caption{The sensing \ac{mi} versus receive SNR.}
	\label{MI vs SNR}
\end{figure}
In Fig. \ref{MI vs SNR}, we show the sensing \ac{mi} versus the received SNR of the echo signal for the various beamforming schemes.
The numbers of communication users and sensing targets are $C=2$ and $K=2$, respectively. 
We observe that the Radar-only approach achieves the highest sensing \ac{mi} among the compared schemes since it is not constrained by communication performance.
Subsequently, the Sensing-centric scheme achieves sensing \ac{mi} second only to Radar-only case, as it focuses solely on meeting the minimum communication performance requirements, with the remaining transmit power allocated towards enhancing sensing capabilities.
The proposed method exhibits superior sensing \ac{mi} performance over the MI-constrained and Communication-centric schemes.
It is also noted that the sensing \ac{mi} of the Communication-centric scheme increases very slowly with SNR, as this scheme solely focuses on meeting the minimum sensing \ac{mi} requirement, allocating the remaining transmit power to improve the communication SINR for each user.
Additionally, we observe that the proposed method closely approaches the ZF-violated scheme.
This indicates that the impact of cross-correlation constraints (\ref{moop2_miup_constraint}) on the achievable maximum sensing \ac{mi}    is minimal.
Nevertheless, in order to fully compare ZF-violated scheme with the proposed method to demonstrate the impact of cross-correlation constraints on overall performance, it is also necessary to consider communication performance.

\begin{figure} [t]
	\centering
	\includegraphics[width=0.40 \textwidth]{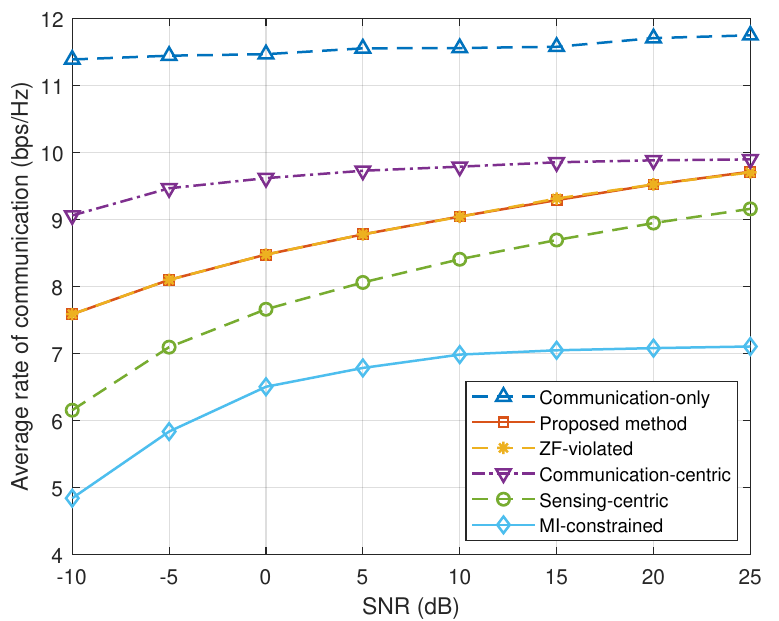}
	\caption{The average rate of the \ac{mu-mimo} communication versus SNR.}
	\label{SINR vs SNR}
\end{figure}
Fig. \ref{SINR vs SNR} unfolds the average achievable communication rate versus communication SNR under the same settings as in Fig. \ref{MI vs SNR}.
It is observed that the Communication-only scheme achieves the best average communication rate, while other schemes experience communication performance loss due to sensing \ac{mi} constraints.
The propose method demonstrates superior average communication rate compared to the \ac{mi}-constrained scheme.
This is due to the fact that the MI-constrained approach optimizes the lower bound of SINR, which cannot attain the same communication performance as the proposed
method that directly optimizes \ac{sinr}s.
Moreover, the performance gap between the proposed method and the Communication-centric approach diminishes as SNR increases, converging closely at high SNR levels.
We observe that the average communication rate of the proposed method closely approximates that of the ZF-violated scheme, suggesting that cross-correlation constraints do not compromise the communication performance either.

\begin{figure} [t]
	\centering
	\includegraphics[width=0.40 \textwidth]{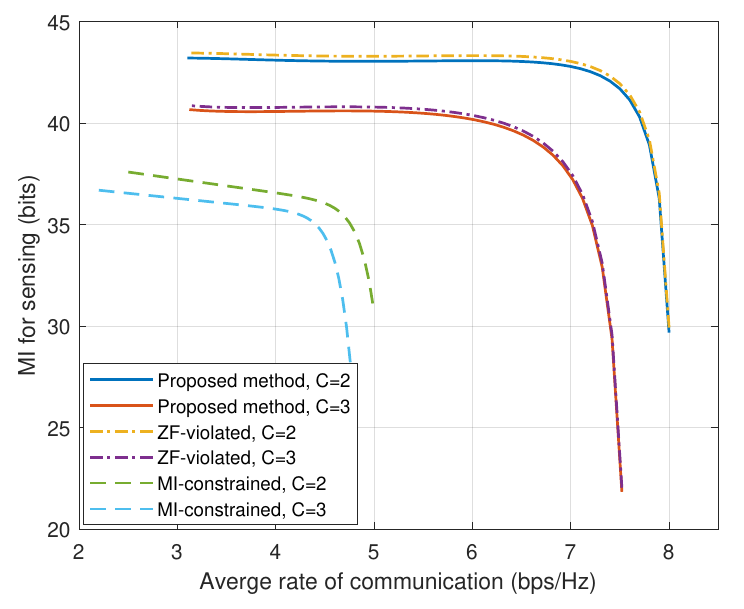}
	\caption{The tradeoff between the sensing \ac{mi} and the average rate of the \ac{mu-mimo} communication.}
	\label{Pareto boundary}
\end{figure}
In  Fig. \ref{Pareto boundary}, we evaluate the tradeoff between the average communication rate and the sensing \ac{mi} for $C=2$ and $C=3$.
For the proposed method, the weight of sensing \ac{mi} $\alpha$ varies from $0.1$ to $0.9$. 
The weights of different communication users are set with the same value, calculated by  ${\omega }_{i}=(1-\alpha)/C, \forall i \in \mathcal{C}$.
The weights of different sensing targets are equivalently established as  ${\xi}_k=\alpha/K, \forall k$.
The threshold $\rho$ utilized in the \ac{mi}-constrained approach varies from $0.2$ to $0.9$.
In the case where the number of users and targets are the same, the proposed method achieves higher sensing \ac{mi} than the \ac{mi}-constrained scheme with the same communication average rate.
Similarly, when to achieve the same sensing \ac{mi}, the proposed method obtains a better average communication rate.
This indicates that the proposed method provides a better trade-off between multi-target sensing performance and multi-user communication performance compared to the MI-constrained scheme.
It is also noted that the proposed method and the ZF-violated scheme  exhibit a similar trade-off,  demonstrating that the cross-correlation constraints does not compromise the overall performance of the \ac{isac} system.
Furthermore, we observe that the more users the \ac{isac} system has to communicate with reliably, the lower the sensing \ac{mi} is achieved in the proposed method.
This is because ensuring an additional communication user performance requires utilizing transmit power resources originally allocated for sensing.
% again indicating the inherent tradeoff between sensing and communications in \ac{isac} systems.
%

\begin{figure} [t!]
	\centering
	\includegraphics[width=0.48 \textwidth]{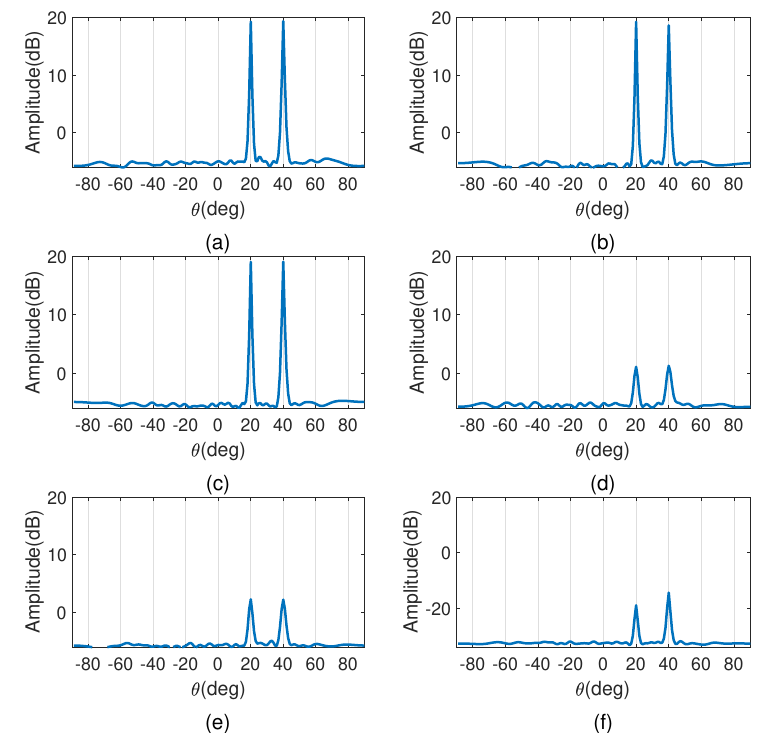}
	\caption{Capon spatial spectrum, for $K=2$. (a) Capon spatial spectrum for the Radar-only approach without radar signals. (b) Capon spatial spectrum for the proposed method. (c) Capon spatial spectrum for the proposed method without  radar signals. (d) Capon spatial spectrum for the ZF-violated scheme. (e) Capon spatial spectrum for the ZF-violated scheme without radar signals. (f) Capon spatial spectrum for the MI-contrained scheme.}
	\label{capon_spectrumK2}
\end{figure}
Since the sensing \ac{mi} is not the only performance measure for sensing, we also examine the angle estimation performance obtained by using the Capon method.
 We simulate two radar targets located at directions $20^{\circ}$ and $40^{\circ}$, respectively.
The complex amplitude of the targets are both $1$\cite{liu2020joint}.
Fig.~\ref{capon_spectrumK2} exhibits the Capon spatial spectrum with and without transmitting radar signals for several benchmarks in one test.
Without transmitting radar signals, it can be observed from Fig~\ref{capon_spectrumK2} (a), (c), (e), and (f) that the proposed method exhibits an angular resolution close to that of the Radar-only approach and yields significantly higher peak values at the desired target directions compared to other ZF-violated and MI-constrained schemes.
We also observe that when the number of targets does not exceed the number of users, all schemes can  effectively detect and estimate the targets regardless of whether additional radar signals are transmitted.

\begin{figure} [t!]
	\centering
	\includegraphics[width=0.47 \textwidth]{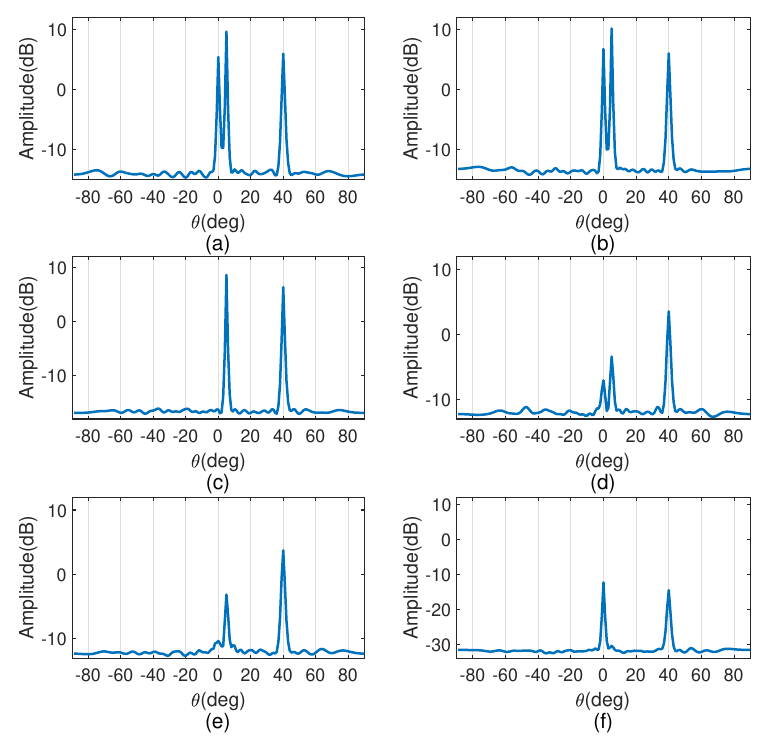}
	\caption{Capon spatial spectrum, for $K=3$. (a) Capon spatial spectrum for the Radar-only approach. (b) Capon spatial spectrum for the proposed method. (c) Capon spatial spectrum for the proposed method without  radar signals. (d) Capon spatial spectrum for the ZF-violated scheme. (e) Capon spatial spectrum for the ZF-violated scheme without radar signals. (f) Capon spatial spectrum for the MI-contrained scheme.}
	\label{capon_spectrumK3}
\end{figure}
Then, we consider the scenario where there are three targets, with two of them placed closely.
The directions of the targets are $0^{\circ}$, $5^{\circ}$ and $40^{\circ}$, and the complex amplitude for each target is $1$\cite{liu2020joint}.
When transmitting radar signals, it can be observed  that the peak at the location of $0^{\circ}$ in Fig~\ref{capon_spectrumK3} (d) is not prominent compared with Fig~\ref{capon_spectrumK3} (b), resulting in an inadequate resolution of two closely spaced targets.
This observation highlights that the proposed method enhances the ability to resolve multiple targets by suppressing the cross-correlation among signals reflected from different targets through constraints (\ref{moop2_miup_constraint}).
Additionally, it can be observed from Fig~\ref{capon_spectrumK3} (c), (e) and (f) that when the BS transmits only communication signals, all three targets cannot be effectively distinguished, with only two of them exhibiting significant peaks.
This is due to the fact that when only communication signals are transmitted, the rank of the transmitted signal covariance matrix is limited by $C$, which cannot provide sufficient \ac{dof}s to form three beams to cover all three targets.

\begin{figure} [t]
	\centering
	\includegraphics[width=0.40 \textwidth]{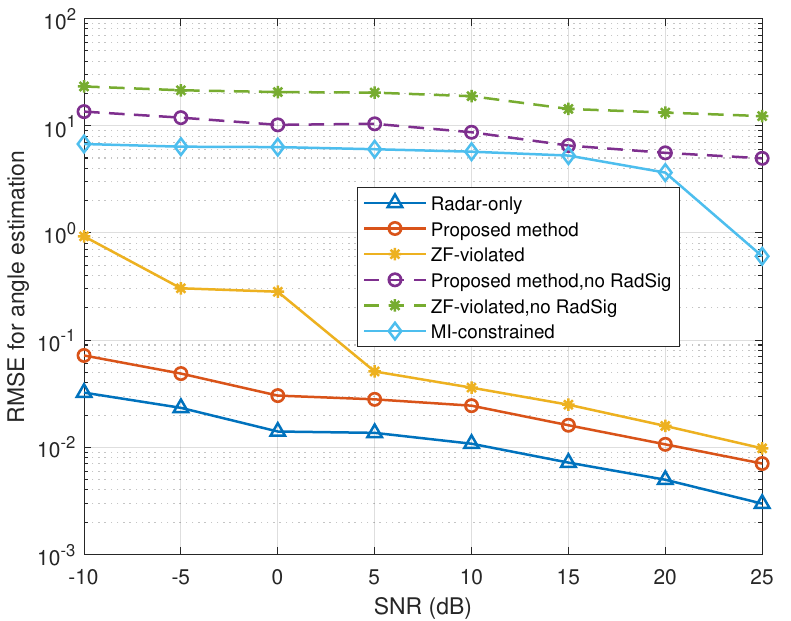}
	\caption{RMSE for angle estimation  versus receive SNR, for $K=3$.}
	\label{angle_rmse}
\end{figure}
 Fig~\ref{angle_rmse}  shows the \ac{rmse} of target angle estimation versus radar \ac{snr}.
 The angles of targets are estimated by finding the peaks in the Capon spatial spectrum.
When radar signals are not transmitted, the \ac{rmse} for the proposed scheme, the ZF-violated scheme, and the MI-constrained scheme is significantly higher compared to other schemes that utilize radar signals.
This is due to the insufficient \ac{dof}s, often resulting in the inability to detect targets close to the true target direction, which aligns with the conclusion in Fig~\ref{capon_spectrumK3}.
Additionally, we observe that when radar signals are transmitted,  the proposed scheme exhibits a lower  \ac{rmse} compared to the ZF-violated scheme  and is close to the  Radar-only scheme. 
This indicates that even with  similar capabilities in maximizing sensing \ac{mi}, the proposed scheme outperforms ZF-violated in terms of angular estimation performance.
Therefore, the  sensing \ac{mi} upper bound with cross-correlation constraints is more appropriate performance metric for multi-target sensing.

\section{Conclusion}
\label{section_conclusion}

In this paper, we investigated a multi-objective optimization framework for \ac{isac} beamforming that provides a flexible tradeoff between multiple targets sensing and multi-user communication.
We formulate a \ac{moop} based on the tight upper bound of sensing \ac{mi} and each user's \ac{sinr}.
 Then, we employ a max-min utility function method to obtain specific Pareto optimal solutions while considering fairness between users and sensing targets.
Numerical results were presented for validating the proposed beamforming method and provided the useful insights.
Firstly, the proposed method can achieve superior performance boundaries of communication and sensing performance, while also facilitating a flexible tradeoff between them.
Secondly, the proposed method can enhance target resolution and angle estimation accuracy for multiple targets, thereby validating the suitability of constrained sensing \ac{mi} upper bound as a performance metric for multi-target sensing.

Extending the proposed framework to practical scenarios with imperfect \ac{csi} may degrade the communication and sensing performance, necessitating the development of robust beamforming techniques that account for channel uncertainties. 
Additionally, dynamic target tracking in non-ideal conditions can be computationally demanding, requiring the development of low-complexity tracking algorithms and efficient beamforming update schemes.
Future research should focus on investigating the robustness of the \ac{mi}-based beamforming method under imperfect CSI and developing advanced techniques to mitigate its impact while enabling real-time adaptation for practical implementation.

\appendices
\section{Proof of Theorem \ref{lemma_mi_trans}} \label{app_mi}
Substituting  (\ref{r_g}) into (\ref{sen_mi_3_22}), the sensing \ac{mi} can be simplified to~(\ref{sen_mi_3}), presented at the top of the next page. 
\begin{figure*}[!t]
    \begin{subequations}\label{sen_mi_3}
		\begin{align}
		\label{sen_mi_3_2}
			I\left( {{{\mathbf{\tilde{y}}}}_{r}};\mathbf{\tilde{g}}\left| {\mathbf{\tilde{X}}} \right. \right) 
			 &= \log \left[ \det \left( \mathbf{I}+\frac{1}{\sigma _{r}^{2}}\left( \sum\limits_{t=1}^{K}{\sigma _{k}^{2}\left( {{\mathbf{X}}^{*}}{{\mathbf{X}}^{T}}\otimes {{\mathbf{I}}_{{{N}_{r}}}} \right)\text{vec}\left( \mathbf{A}\left( {{\theta }_{k}} \right) \right)}\text{vec}{{\left( \mathbf{A}\left( {{\theta }_{k}} \right) \right)}^{H}} \right) \right) \right] \\ 
			 \label{sen_mi_3_3}
			& =\log \left[ \det \left( \mathbf{I}+\frac{1}{\sigma _{r}^{2}}\left( \sum\limits_{t=1}^{K}{\sigma _{k}^{2}\text{vec}\left( \mathbf{A}\left( {{\theta }_{k}} \right)\left( \mathbf{X}{{\mathbf{X}}^{H}} \right) \right)}\text{vec}{{\left( \mathbf{A}\left( {{\theta }_{k}} \right) \right)}^{H}} \right) \right) \right],
			 \end{align}
		\end{subequations}
  \rule[-0pt]{18.5 cm}{0.05em}%\vspace{-0.2 cm}
\end{figure*}
The derivation of (\ref{sen_mi_3_3}) follows the fact that $\text{vec}(\mathbf{ABC})=({{\mathbf{C}}^{T}}\otimes \mathbf{A}) \text{vec}(\mathbf{B})$.

Then, by introducing auxiliary block  matrices
\begin{equation}
	\label{MT_MI}
	\begin{aligned}
		& \mathbf{T}=\frac{1}{\sigma _{r}^{2}}\left[ \text{vec}\left(\! \mathbf{A}\left( {{\theta }_{1}} \right)\!\left( \mathbf{X}{{\mathbf{X}}^{H}} \right) \!\right)\!,\ldots ,\text{vec}\left( \mathbf{A}\left( {{\theta }_{K}}\! \right)\!\left( \mathbf{X}{{\mathbf{X}}^{H}} \right) \right) \right], \\ 
		& \mathbf{M}={{\left[ \sigma _{1}^{2}\text{vec}\left( \mathbf{A}\left( {{\theta }_{1}} \right) \right),\ldots ,\sigma _{K}^{2}\text{vec}\left( \mathbf{A}\left( {{\theta }_{K}} \right) \right) \right]}^{H}},  
	\end{aligned}
\end{equation}
the sensing \ac{mi} in (\ref{sen_mi_3_3}) can be rewritten as
\begin{equation}
	\label{sen_mi_mt}
	\begin{aligned}
		I\left( {{{\mathbf{\tilde{y}}}}_{r}};\mathbf{\tilde{g}}\left| {\mathbf{\tilde{X}}} \right. \right)& =\log \left[ \det \left( \mathbf{I}+\mathbf{TM} \right) \right] 
        =\log \left[ \det \left( \mathbf{I}+\mathbf{MT} \right) \right], 		 	\end{aligned}
\end{equation}
where the last equality follows from Sylvester's determinant identity $\det (\mathbf{I}+\mathbf{AB})=\det (\mathbf{I}+\mathbf{BA})$. The $(i,j)$-th entry of $\mathbf{MT}$ is given by
\begin{equation}
	\begin{aligned}\label{MT2}
		 {{\left[ \mathbf{MT} \right]}_{i,j}}& =\frac{\sigma _{i}^{2}}{\sigma _{r}^{2}}\text{ve}{{\text{c}}^{H}}\left( \mathbf{A}\left( {{\theta }_{i}} \right) \right)\text{vec}\left( \mathbf{A}\left( {{\theta }_{j}} \right)\left( \mathbf{X}{{\mathbf{X}}^{H}} \right) \right) \\
		%& =\frac{\sigma _{i}^{2}}{\sigma _{r}^{2}}\text{tr}\left( {{\mathbf{A}}^{H}}\left( {{\theta }_{i}} \right)\mathbf{A}\left( {{\theta }_{j}} \right)\mathbf{X}{{\mathbf{X}}^{H}} \right) 	\label{MT1} \\ 
		 &=\frac{L\sigma _{i}^{2}}{\sigma _{r}^{2}}\text{tr}\left( {{\mathbf{A}}^{H}}\left( {{\theta }_{i}} \right)\mathbf{A}\left( {{\theta }_{j}} \right){{\mathbf{R}}_{X}} \right)=\sigma _{i}^{2} {{\left[ \mathbf{\Phi} \right]}_{i,j}},		 
	\end{aligned}
\end{equation}
where the derivation of (\ref{MT2}) utilizes the identity $\text{ve}{{\text{c}}^{H}}\left( \mathbf{A} \right)\text{vec}\left( \mathbf{B} \right)=\text{tr}\left( {{\mathbf{A}}^{H}}\mathbf{B} \right)$, and ${{\left[ \mathbf{\Phi} \right]}_{i,j}}$ denotes the $(i,j)$-th element of the matrix $\mathbf{\Phi} \in \mathbb{C}^{K \times K}$.
Next, we define a diagonal matrix $\mathbf{\Lambda}\in \mathbb{C}^{K \times K}$ as
\begin{equation}
\label{lamda}
\mathbf{\Lambda} \triangleq \text{diag}\left\{ \frac{1}{\sigma _{1}^{2}},\frac{1}{\sigma _{2}^{2}},\ldots ,\frac{1}{\sigma _{K}^{2}} \right\}.
\end{equation}
By extracting the constant coefficient terms $\{\sigma _{1}^{2},\ldots, \sigma _{K}^{2}\}$ from determinant by row, the sensing \ac{mi} is recast as 
\begin{equation}
	\label{sen_mi_ph}
	\begin{aligned}
		I\left( {{{\mathbf{\tilde{y}}}}_{r}};\mathbf{\tilde{g}}\left| {\mathbf{\tilde{X}}} \right. \right) &=\log \left[  \det \left( \mathbf{\Phi }+\mathbf{\Lambda } \right)  \prod \limits_{k=1}^K\sigma _{k}^{2}\right],\\
	\end{aligned}
\end{equation}
which completes the proof of Lemma~\ref{lemma_mi_trans}.

\section{Proof of Lemma \ref{lemma_1}}\label{appc}
We first prove that $\mathcal{M}$ is a compact set.
Following \cite{bjornson2012robust}, we know that the feasible transmit strategy set $\mathcal{R}$ is compact.
According to \cite{rudin1976principles}, the continuous mapping of a compact set is also a compact set.
Since the upper bound of sensing \ac{mi} $I_{\rm up}$ and each user's  \ac{sinr} ${\gamma }_{i}$ are all continuous functions of $\left\{ {{\mathbf{R}}_{1}},\ldots ,{{\mathbf{R}}_{{{C}}+M}}\right\} \in {\mathcal{R}}$,  $\mathcal{M}$ is a compact set.

To prove that $\mathcal{M}$ is a normal set, we use a similar method as described in  \cite[Appx.A]{chen2021joint}. 
Take $\mathbf{x}=\left( {{I}_{\rm up}},{{\gamma }_{1}},\cdots ,{{\gamma }_{{{C}}}} \right)\in \mathcal{M}$ and assume that $\left( {{\mathbf{R}}_{1}},\ldots ,{{\mathbf{R}}_{{{C}}+M}} \right)$ is a feasible strategy that attains point $\mathbf{x}$.
Our goal is to prove that any given $\mathbf{x}^{\prime}=\left( {I^{\prime}}_{\rm up}, {\gamma^{\prime}}_{1},\cdots, {\gamma^{\prime}}_{C} \right)$ satisfying $\mathbf{x}^{\prime}\le \mathbf{x}$ also belongs to $\mathcal{M}$.

To this end, we represent the transmit beamforming strategy in a new form $\left( {{p}_{1}}{{\mathbf{R}}_{1}},\ldots ,{{p}_{{{C}}+M}}{{\mathbf{R}}_{{{C}}+M}} \right)$, where $\left( {{p}_{1}},\ldots ,{{p}_{{{C}}+M}} \right)\in \mathcal{P}$ is a set of power allocation coefficients that should belong to
\begin{equation}
	\label{Power_coef}
	\begin{aligned}
	\mathcal{P} = \{ & \left( {{p}_{1}},\ldots,{{p}_{{{C}}+M}} \right) : p_n > 0, \text{tr}\left(\sum\nolimits_{n=1}^{{{C}}+M}{{{p}_{n}}{{\mathbf{R}}_{n}}} \right) \le {{P}_{T}}, \\
	& \text{tr}\left(\mathbf{a}\left( {{\theta }_{i}} \right){{\mathbf{a}}^{H}}\left( {{\theta }_{j}} \right)\left(\sum\nolimits_{n=1}^{{{C}}+M}{{{p}_{n}}{{\mathbf{R}}_{n}}} \right) \right)=0 \quad (i \ne j) \}.
	\end{aligned}
\end{equation}
First, we need to find that $\left( {{p}_{1}},\ldots ,{{p}_{{{C}}+M}} \right)\in \mathcal{P}$ satisfies the given \ac{sinr}s and $I_{\rm up}$, i.e., 
\begin{subequations}
	\begin{align}
		& {{\gamma }_{i}}=\frac{{{p}_{i}}\mathbf{h}_{i}^{H}{{\mathbf{R}}_{i}}{{\mathbf{h}}_{i}}}{\sum\nolimits_{j=1,j\ne i }^{{{C}}+M}{{{p}_{j}}\mathbf{h}_{i}^{H}{{\mathbf{R}}_{j}}{{\mathbf{h}}_{i}}}+\sigma _{c}^{2}}, \forall i \in \mathcal{C},
		 	\label{p_sinr}\\ 
		& {{I}_{\rm up}}=\sum\limits_{{k}=1}^{K}{\log\! \left( \!\!1+\! \frac{{{N}_{r}}\sigma _{k}^{2}L}{\sigma _{r}^{2}}{{\mathbf{a}}^{H}}\!\!\left( {{\theta }_{k}} \right)\!\left(\! \sum\limits_{n=1}^{C+M}{p_n}{{\mathbf{R}}_{n}} \right)\!\mathbf{a}\left( {{\theta }_{k}} \right) \!\!\right)}.\label{p_mi}		 
	\end{align}
\end{subequations}

For (\ref{p_sinr}), given arbitrary $\left( {{p}_{C+1}},\ldots ,{{p}_{{{C}}+M}} \right)$ satisfying each element $p_i \in(0,1]$, we can derive the following equation, i.e.,
\begin{equation}
	\label{prove_sinr_eq1}
	\begin{aligned}
		\mathbf{\Gamma }\left( {\mathbf{A}_1} {\mathbf{p}_1}+ {\mathbf{A}_2} {\mathbf{p}_2} + \mathbf{b}\right)=\mathbf{p}_1,
	\end{aligned}
\end{equation}
where $\mathbf{\Gamma }=\text{diag}\left( {{\gamma }_{i}},\ldots ,{{\gamma }_{{{C}}}} \right)$,  $\mathbf{p}_1=\!{{\left[ {{p}_{1}},\ldots, {{p}_{{{C}}}} \right]}^{T}}$, $\mathbf{p}_2=\!{{\left[ {{p}_{C+1}},\ldots,  {{p}_{C+M}} \right]}^{T}}$,
\begin{equation}
	\label{A_sinr}
	\begin{aligned}
     \mathbf{A}_1= \left[ \begin{matrix}
        0 & a_{1,2} & \cdots & a_{1,C} \\
        a_{2,1} & 0 & \cdots & a_{2,C} \\
       \vdots & \vdots & \ddots & \vdots \\
          a_{C,1} & a_{C,2} & \cdots & 0
        \end{matrix} \right], 
	\end{aligned}
\end{equation}
\begin{equation}
    \begin{aligned}
         \mathbf{A}_2=\left[ \begin{matrix}
			a_{1,C+1} & \cdots & a_{1,C+M} \\
                a_{2,C+1} & \cdots & a_{2,C+M} \\
                \vdots & \ddots & \vdots \\
                a_{C,C+1} & \cdots & a_{C,C+M}
		\end{matrix} \right],
    \end{aligned}
\end{equation}
with
\begin{equation}
	\label{a_sinr}
	\begin{aligned}
		{{a}_{i,j}}=\frac{\mathbf{h}_{i}^{H}{{\mathbf{R}}_{j}}{{\mathbf{h}}_{i}}}{\mathbf{h}_{i}^{H}{{\mathbf{R}}_{i}}{{\mathbf{h}}_{i}}},			 		 
	\end{aligned}
\end{equation}
and $\mathbf{b}={{\left[ {{b}_{1}},\ldots ,{{b}_{{{C}}}} \right]}^{T}}\in {{\mathbb{R}}^{{{C}}\times 1}}$ with
\begin{equation}
	\label{b_sinr}
	\begin{aligned}
		{{b}_{i}}=\frac{\sigma _{c}^{2}}{\mathbf{h}_{i}^{H}{{\mathbf{R}}_{i}}{{\mathbf{h}}_{i}}}.				 
	\end{aligned}
\end{equation}
Based on (\ref{prove_sinr_eq1}),  we can obtain that
\begin{equation}
	\label{P_sinr}
	\begin{aligned}	
		\mathbf{p}_1& ={{\left( \mathbf{I}-{\mathbf{\Gamma }}{{\mathbf{A }}_1} \right)}^{-1}}{\left({\mathbf{\Gamma }}{\mathbf{A}_2}{\mathbf{p}_2}+{\mathbf{\Gamma }}\mathbf{b}\right)}\\ 
		& \overset{\left(a\right)}{=}\left( \mathbf{I}+\sum\limits_{n=1}^{\infty }{{{\left( {\mathbf{\Gamma }}{{\mathbf{A }}_1} \right)}^{n}}} \right){\left({\mathbf{\Gamma }}{\mathbf{A}_2}{\mathbf{p}_2}+{\mathbf{\Gamma }}\mathbf{b}\right)},   
	\end{aligned}
\end{equation}
where $(a)$ is based on the  Neumann series approximation\cite{stewart1998matrix} of ${{\left( \mathbf{I}-{\mathbf{\Gamma }}{{\mathbf{A }}_1} \right)}^{-1}}$.
Since all elements in ${{\mathbf{A }}_1}$, ${{\mathbf{A }}_2}$, $\mathbf{p}_2$ and $\mathbf{b}$ are non-negative, it indicates that each element in $\mathbf{p}_1$ is obtained from a polynomial with positive coefficients in ${{\gamma }_{i}},\ldots ,{{\gamma }_{{{C}}}}$.
Therefore, a decrease in $\gamma_{i}$  will necessarily lead to a decrease in at least one of the power allocation coefficients $\left( {{p}_{1}},\ldots, {{p}_{{{C}}}} \right)$.
For (\ref{p_mi}),  $I_{\rm up}$ can be  further rewritten as
\begin{equation}\label{iup_p}
    {{I}_{\rm up}}=\log \left( \prod\nolimits_{k=1}^{K}{\left( 1+\sum\nolimits_{n=1}^{C+M}{{{p}_{n}}{{d}_{nk}}} \right)} \right),
\end{equation}
where ${d}_{nk}$ is a positive constant given as
\begin{equation}
   {d}_{nk}=  \frac{{{N}_{r}}\sigma _{k}^{2}L}{\sigma _{r}^{2}}{{\mathbf{a}}^{H}}\left( {{\theta }_{k}} \right) {\mathbf{R}}_{n}\mathbf{a}\left( {{\theta }_{k}} \right), \forall n,k.
\end{equation}
From (\ref{iup_p}), it is evident that $I_{\rm up}$ can be expressed as the logarithm of a polynomial in ${{p}_{1}},\ldots, {p}_{C+M}$ with positive coefficients.
As the logarithm function is monotonically increasing, a reduction in $I_{\rm up}$ will inevitably result in a decrease in at least one of the power allocation coefficients $\left( {{p}_{1}},\ldots, {{p}_{C+M}} \right)$.

In summary,  based on the above analysis, we can conclude that for any ${\gamma^{\prime}}_{i}\le {{\gamma }_{i}}, \forall i$ or ${I^{\prime}}_{\rm up}\le {{I}_{\rm up}}$, we can always properly find  $\mathbf{p}^{\prime}\in \mathcal{P}$ satisfying that $\mathbf{p}^{\prime}\le\mathbf{p}$, and making $\left( {{p}^{\prime}_{1}}{{\mathbf{R}}_{1}},\ldots ,{{p}^{\prime}_{{{C}}+M}}{{\mathbf{R}}_{{{C}}+M}} \right)$ a feasible solution that attains point $\mathbf{x}^{\prime}$.
This implies that if $\mathbf{x}^{\prime}\in \mathbb{R}_{+}^{n}$ and satisfies $\mathbf{x}^{\prime}\le \mathbf{x}$, then $\mathbf{x}^{\prime}$ belongs to $\mathcal{M}$.
In other words, the achievable performance region $\mathcal{M}$ is a normal set, which completes the proof. 

\vspace{-0.5cm}
\section{Proof of Theorem \ref{lemma_2}} \label{appd}
The optimal values of (\ref{moop_4}) is denoted as $\left( I_{\rm up}^{*},\gamma _{1}^{*},\cdots ,\gamma _{C}^{*} \right)$, and the corresponding optimal solution with arbitrary ranks as 
\begin{equation}
	\label{opt_arbrank}
	\begin{aligned}		
		 \left\{ \mathbf{R}_{1}^{*},\ldots ,\mathbf{R}_{{C}+M}^{*} \right\}.
	\end{aligned}
\end{equation}

We need to prove that the rank-one optimal solution $\left\{ {{{\mathbf{\bar{R}}}}_{1}},\ldots ,{{{\mathbf{\bar{R}}}}_{{C}+M}} \right\}$ can be constructed from (\ref{opt_arbrank}).
First, we construct  $\left\{ {{{\mathbf{\bar{R}}}}_{1}},\ldots ,{{{\mathbf{\bar{R}}}}_{C}} \right\}$ as
\begin{equation}
	\label{opt_ran1_com}
	\begin{aligned}		
		{{\mathbf{\bar{R}}}_{i}}={{\mathbf{\bar{w}}}_{i}}\mathbf{\bar{w}}_{i}^{H}, \forall i \in \mathcal{C},
	\end{aligned}
\end{equation}
where
\begin{equation}
	\label{opt_ran1_bf}
	\begin{aligned}		
		{{\mathbf{\bar{w}}}_{i}}=\frac{\mathbf{R}_{i}^{*}{{\mathbf{h}}_{i}}}{\sqrt{\mathbf{h}_{i}^{H}\mathbf{R}_{i}^{*}{{\mathbf{h}}_{i}}}} , \forall i \in \mathcal{C},
	\end{aligned}
\end{equation}
It can be readily verified that
\begin{equation}
	\label{opt_ran1_bf1}
	\begin{aligned}		
		\mathbf{h}_{i}^{H}{{\mathbf{\bar{R}}}_{i}}{{\mathbf{h}}_{i}}=\mathbf{h}_{i}^{H}{{\mathbf{\bar{w}}}_{i}}\mathbf{\bar{w}}_{i}^{H}\mathbf{h}=\mathbf{h}_{i}^{H}\mathbf{R}_{i}^{*}{{\mathbf{h}}_{i}}, \forall i \in \mathcal{C},
	\end{aligned}
\end{equation}
By noting the above fact, we next construct  $\left\{ {{{\mathbf{\bar{R}}}}_{C+1}},\ldots ,{{{\mathbf{\bar{R}}}}_{C+M}}\right\}$. 
Let ${{\mathbf{R}}_{r}}\triangleq\sum\nolimits_{n=1}^{{C}+M}{\mathbf{R}_{n}^{*}}-\sum\nolimits_{n=1}^{{C}}{{{{\mathbf{\bar{R}}}}_{n}}}=\sum\nolimits_{n=C+1}^{C+M}{{{{\mathbf{\bar{R}}}}_{n}}}$.
Therefore,  we need to prove that ${{\mathbf{R}}_{r}}$ is semidefinite.
Since  $\sum\nolimits_{n=C+1}^{{C}+M}{\mathbf{R}_{n}^{*}}$ is semidefinite due to (\ref{semi_constraint}), we only need to prove that $\mathbf{R}_{i}^{*}-{{\mathbf{\bar{R}}}_{i}}\succeq 0$ for all $i\leqslant C$.

For any $\mathbf{x}\in {{\mathbb{C}}^{N_t}}$, it holds that 
\begin{equation}
	\label{proof_Rr_1}
	\begin{aligned}		
		{{\mathbf{x}}^{H}}\left( \mathbf{R}_{i}^{*}-{{{\mathbf{\bar{R}}}}_{i}} \right)\mathbf{x}={{\mathbf{x}}^{H}}\mathbf{R}_{i}^{*}\mathbf{x}-\frac{{{\left| {{\mathbf{x}}^{H}}\mathbf{R}_{i}^{*}{{\mathbf{h}}_{i}} \right|}^{2}}}{\left( \mathbf{h}_{i}^{H}\mathbf{R}_{i}^{*}{{\mathbf{h}}_{i}} \right)}.
	\end{aligned}
\end{equation}

According to Cauchy-Schwarz inequality, we have
\begin{equation}
	\label{proof_Rr_2}
	\begin{aligned}		
		\left( \mathbf{h}_{i}^{H}\mathbf{R}_{i}^{*}{{\mathbf{h}}_{i}} \right)\left( {{\mathbf{x}}^{H}}\mathbf{R}_{i}^{*}\mathbf{x} \right)\ge {{\left| {{\mathbf{x}}^{H}}\mathbf{R}_{i}^{*}{{\mathbf{h}}_{i}} \right|}^{2}},
	\end{aligned}
\end{equation}
which means that 
\begin{equation}
	\label{proof_Rr_3}
	\begin{aligned}		
		  {{\mathbf{x}}^{H}}\left( \mathbf{R}_{i}^{*}-{{{\mathbf{\bar{R}}}}_{i}} \right)\mathbf{x}\ge 0 ,\ \forall \mathbf{x}\in {{\mathbb{C}}^{N}} ,
	\end{aligned}
\end{equation}
i.e., ${{\mathbf{R}}_{r}}\succeq 0$.
 
By taking the Cholesky decomposition of ${{\mathbf{R}}_{r}}$, we have 
\begin{equation}
	\label{proof_Rr_4}
	\begin{aligned}		
{{\mathbf{\bar{W}}}_{r}}\mathbf{\bar{W}}_{r}^{H}=\sum\nolimits_{n=1}^{C+M}{\mathbf{R}_{n}^{*}}-\sum\nolimits_{n=1}^{C}{{{{\mathbf{\bar{R}}}}_{n}}},
	\end{aligned}
\end{equation}
where ${{\mathbf{\bar{W}}}_{r}}=\left[ {{{\mathbf{\bar{w}}}}_{{C}+1}},\ldots ,{{{\mathbf{\bar{w}}}}_{{C}+M}} \right]$ is a lower triangular matrix.
So we obtain that ${{\mathbf{\bar{R}}}_{j}}={{\mathbf{\bar{w}}}_{j}}\mathbf{\bar{w}}_{j}^{H}$ for $j=C+1,\ldots ,C+M$.
It follows that
\begin{equation}
	\label{proof_ob_1_1}
	\begin{aligned}		
		\sum\nolimits_{n=1}^{{C}+M}{\mathbf{R}_{n}^{*}}=\sum\nolimits_{n=1}^{{C}+M}{{{{\mathbf{\bar{R}}}}_{n}}}.
	\end{aligned}
\end{equation} 

Now we need to validate that $\left\{ {{{\mathbf{\bar{R}}}}_{1}},\ldots ,{{{\mathbf{\bar{R}}}}_{{C}+M}} \right\}$ is  a feasible solution to (\ref{moop_4}).
 And we have
\begin{equation}
	\label{proof_cons1}
	\begin{aligned}		
		&\text{tr}\left( \sum\nolimits_{n=1}^{C+M}{{{{\mathbf{\bar{R}}}}_{n}}} \right)=\text{ tr}\left( \sum\nolimits_{n=1}^{C+M}{\mathbf{R}_{n}^{*}} \right)\le {{P}_{T}},\\
	& \text{tr}\left( \mathbf{a}\left( {{\theta }_{i}} \right){{\mathbf{a}}^{H}}\left( {{\theta }_{j}} \right)\left( \sum\nolimits_{n=1}^{{C}+M}{{{{\mathbf{\bar{R}}}}_{n}}} \right) \right) = \\ &\text{tr}\left( \mathbf{a}\left( {{\theta }_{i}} \right){{\mathbf{a}}^{H}}\left( {{\theta }_{j}} \right)\left(\sum\nolimits_{n=1}^{{C}+M}{\mathbf{R}_{n}^{*}} \right) \right)\!=\!0 \left( i\ne j \right),
	\end{aligned}
\end{equation} 
which means that the constraints of (\ref{moop_4}) hold for $\{ {{{\mathbf{\bar{R}}}}_{1}},\ldots ,{{{\mathbf{\bar{R}}}}_{{C}+M}}\}$.

Based on (\ref{opt_ran1_bf1}), we have
\begin{equation}
	\label{proof_obj1}
	\begin{aligned}		
		{I_{\rm up}^{*}}& =\sum\nolimits_{{k}=1}^{K}{\log\! \left( 1+\! \frac{{{N}_{r}}\sigma _{k}^{2}L}{\sigma _{r}^{2}}{{\mathbf{a}}^{H}}\!\!\left( {{\theta }_{k}} \right)\!\left(\! \sum\limits_{n=1}^{C+M}{\mathbf{R}_{n}^{*}} \right)\!\mathbf{a}\left( {{\theta }_{k}} \right) \right)}\\ 
		& =\sum\nolimits_{{k}=1}^{K}{\log\! \left( 1+\! \frac{{{N}_{r}}\sigma _{k}^{2}L}{\sigma _{r}^{2}}{{\mathbf{a}}^{H}}\!\!\left( {{\theta }_{k}} \right)\!\left(\! \sum\limits_{n=1}^{C+M}{{{{\mathbf{\bar{R}}}}_{n}}}\right)\!\mathbf{a}\left( {{\theta }_{k}} \right) \right)},
	\end{aligned}
\end{equation}
\begin{equation}
	\label{proof_obj2}
	\begin{aligned}		
			{\gamma _{i}^{*}}	& =\frac{\mathbf{h}_{i}^{H}\mathbf{R}_{i}^{*}{{\mathbf{h}}_{i}}}{\mathbf{h}_{i}^{H}\left( \sum\nolimits_{n=1}^{{C}+M}{\mathbf{R}_{n}^{*}} \right){{\mathbf{h}}_{i}}-\mathbf{h}_{i}^{H}\mathbf{R}_{i}^{*}{{\mathbf{h}}_{i}}+\sigma _{c}^{2}} \\
   &=\frac{\mathbf{h}_{i}^{H}{{{\mathbf{\bar{R}}}}_{i}}{{\mathbf{h}}_{i}}}{\mathbf{h}_{i}^{H}\left( \sum\nolimits_{n=1}^{{C}+M}{{{{\mathbf{\bar{R}}}}_{n}}} \right){{\mathbf{h}}_{i}}-\mathbf{h}_{i}^{H}{{{\mathbf{\bar{R}}}}_{i}}{{\mathbf{h}}_{i}}+\sigma _{c}^{2}}, 
	\end{aligned}
\end{equation}
which means that the optimal values do not change.

Above all, $\{{{{\mathbf{\bar{R}}}}_{1}},\ldots ,{{{\mathbf{\bar{R}}}}_{{C}+M}}\}$ is also the optimal solution to \eqref{moop_4}, which completes the proof.

\bibliographystyle{IEEEtran}
\bibliography{mybib}

\begin{thebibliography}{10}
\providecommand{\url}[1]{#1}
\csname url@samestyle\endcsname
\providecommand{\newblock}{\relax}
\providecommand{\bibinfo}[2]{#2}
\providecommand{\BIBentrySTDinterwordspacing}{\spaceskip=0pt\relax}
\providecommand{\BIBentryALTinterwordstretchfactor}{4}
\providecommand{\BIBentryALTinterwordspacing}{\spaceskip=\fontdimen2\font plus
\BIBentryALTinterwordstretchfactor\fontdimen3\font minus
  \fontdimen4\font\relax}
\providecommand{\BIBforeignlanguage}[2]{{%
\expandafter\ifx\csname l@#1\endcsname\relax
\typeout{** WARNING: IEEEtran.bst: No hyphenation pattern has been}%
\typeout{** loaded for the language `#1'. Using the pattern for}%
\typeout{** the default language instead.}%
\else
\language=\csname l@#1\endcsname
\fi
#2}}
\providecommand{\BIBdecl}{\relax}
\BIBdecl

\bibitem{liu2020state}
F.~Liu, C.~Masouros, A.~P. Petropulu, H.~Griffiths, and L.~Hanzo, ``Joint radar
  and communication design: Applications, state-of-the-art, and the road
  ahead,'' \emph{IEEE Trans. Commun.}, vol.~68, no.~6, pp. 3834--3862, Feb.
  2020.

\bibitem{liu2022integrated}
F.~Liu, Y.~Cui, C.~Masouros, J.~Xu, T.~X. Han, Y.~C. Eldar, and S.~Buzzi,
  ``Integrated sensing and communications: Toward dual-functional wireless
  networks for {6G} and beyond,'' \emph{{IEEE} J. Select. Areas Commun.},
  vol.~40, no.~6, pp. 1728--1767, Mar. 2022.

\bibitem{ma2020joint}
D.~Ma, N.~Shlezinger, T.~Huang, Y.~Liu, and Y.~C. Eldar, ``Joint
  radar-communication strategies for autonomous vehicles: Combining two key
  automotive technologies,'' \emph{{IEEE} Signal Processing Mag.}, vol.~37,
  no.~4, pp. 85--97, Jun. 2020.

\bibitem{feng2021joint}
Z.~Feng, Z.~Wei, X.~Chen, H.~Yang, Q.~Zhang, and P.~Zhang, ``Joint
  communication, sensing, and computation enabled {6G} intelligent machine
  system,'' \emph{{IEEE} Network}, vol.~35, no.~6, pp. 34--42, Nov. 2021.

\bibitem{li2024uav}
X.~Li, M.~Zhang, H.~Chen, C.~Han, L.~Li, D.-T. Do, S.~Mumtaz, and
  A.~Nallanathan, ``{UAV}-enabled multi-pair massive {MIMO-NOMA} relay systems
  with low-resolution {ADCs/DACs},'' \emph{{IEEE} Trans. Veh. Technol.},
  vol.~73, no.~2, pp. 2171--2186, Feb. 2024.

\bibitem{yuan2024nest}
Y.~Cheng, J.~Du, J.~Liu, L.~Jin, X.~Li, and D.~B. da~Costa, ``Nested
  tensor-based framework for {ISAC} assisted by reconfigurable intelligent
  surface,'' \emph{{IEEE} Trans. Veh. Technol.}, vol.~73, no.~3, pp.
  4412--4417, Mar. 2024.

\bibitem{zhang2021overview}
J.~A. Zhang, F.~Liu, C.~Masouros, R.~W. Heath, Z.~Feng, L.~Zheng, and
  A.~Petropulu, ``An overview of signal processing techniques for joint
  communication and radar sensing,'' \emph{{IEEE} J. Sel. Top. Sign. Proces.},
  vol.~15, no.~6, pp. 1295--1315, Sept. 2021.

\bibitem{zhang2022enabling}
J.~A. Zhang, M.~L. Rahman, K.~Wu, X.~Huang, Y.~J. Guo, S.~Chen, and J.~Yuan,
  ``Enabling joint communication and radar sensing in mobile networks—a
  survey,'' \emph{{IEEE} Commun. Surv. Tutor.}, vol.~24, no.~1, pp. 306--345,
  Oct. 2022.

\bibitem{fang2023joint}
X.~Fang, W.~Feng, Y.~Chen, N.~Ge, and Y.~Zhang, ``Joint communication and
  sensing toward {6G}: Models and potential of using {MIMO},'' \emph{IEEE
  Internet of Things Journal}, vol.~10, no.~5, pp. 4093--4116, 2023.

\bibitem{wang2023nearisac}
Z.~Wang, X.~Mu, and Y.~Liu, ``Near-field integrated sensing and
  communications,'' \emph{IEEE Commun. Lett.}, vol.~27, no.~8, pp. 2048--2052,
  Aug. 2023.

\bibitem{liu2018mumimo}
F.~Liu, C.~Masouros, A.~Li, H.~Sun, and L.~Hanzo, ``{MU-MIMO} communications
  with {MIMO} radar: From co-existence to joint transmission,'' \emph{IEEE
  Transactions on Wireless Communications}, vol.~17, no.~4, pp. 2755--2770,
  2018.

\bibitem{liu2022transmit}
X.~Liu, T.~Huang, and Y.~Liu, ``Transmit design for joint {MIMO} radar and
  multiuser communications with transmit covariance constraint,'' \emph{IEEE J.
  Sel. Areas Commun.}, vol.~40, no.~6, pp. 1932--1950, Mar. 2022.

\bibitem{liu2020joint}
X.~Liu, T.~Huang, N.~Shlezinger, Y.~Liu, J.~Zhou, and Y.~C. Eldar, ``Joint
  transmit beamforming for multiuser {MIMO} communications and {MIMO} radar,''
  \emph{IEEE Trans. Signal Process.}, vol.~68, pp. 3929--3944, Jun. 2020.

\bibitem{chen2021joint}
L.~Chen, F.~Liu, W.~Wang, and C.~Masouros, ``Joint radar-communication
  transmission: A generalized pareto optimization framework,'' \emph{IEEE
  Trans. Signal Process.}, vol.~69, pp. 2752--2765, 2021.

\bibitem{ni2021multi}
Z.~Ni, J.~A. Zhang, K.~Yang, X.~Huang, and T.~A. Tsiftsis, ``Multi-metric
  waveform optimization for multiple-input single-output joint communication
  and radar sensing,'' \emph{IEEE Trans. Commun.}, vol.~70, no.~2, pp.
  1276--1289, Dec. 2022.

\bibitem{XinyuanMI}
X.~Yuan, Z.~Feng, J.~A. Zhang, W.~Ni, R.~P. Liu, Z.~Wei, and C.~Xu,
  ``Spatio-temporal power optimization for {MIMO} joint communication and radio
  sensing systems with training overhead,'' \emph{IEEE Trans. Veh. Technol.},
  vol.~70, no.~1, pp. 514--528, Dec. 2021.

\bibitem{chunwei2022}
C.~Meng, Z.~Wei, and Z.~Feng, ``Adaptive waveform optimization for {MIMO}
  integrated sensing and communication systems based on mutual information,''
  in \emph{2022 14th International Conference on Wireless Communications and
  Signal Processing (WCSP)}, Nov. 2022, pp. 472--477.

\bibitem{liu2021cramer}
F.~Liu, Y.-F. Liu, A.~Li, C.~Masouros, and Y.~C. Eldar, ``Cram{\'e}r-rao bound
  optimization for joint radar-communication beamforming,'' \emph{IEEE Trans.
  Signal Process.}, vol.~70, pp. 240--253, 2022.

\bibitem{Hua2022mimo}
H.~Hua, X.~Song, Y.~Fang, T.~X. Han, and J.~Xu, ``{MIMO} integrated sensing and
  communication with extended target: {CRB}-rate tradeoff,'' in \emph{GLOBECOM
  2022 - 2022 IEEE Global Communications Conference}, Dec. 2022, pp.
  4075--4080.

\bibitem{ren2023fund}
Z.~Ren, Y.~Peng, X.~Song, Y.~Fang, L.~Qiu, L.~Liu, D.~W.~K. Ng, and J.~Xu,
  ``Fundamental crb-rate tradeoff in multi-antenna {ISAC} systems with
  information multicasting and multi-target sensing,'' \emph{{IEEE} Trans.
  Wirel. Commun.}, pp. 1--1, Sept. 2023.

\bibitem{sun2023trade}
J.~Sun, S.~Ma, G.~Xu, and S.~Li, ``Trade-off between positioning and
  communication for millimeter wave systems with {Ziv-Zakai} bound,''
  \emph{IEEE Trans. Commun.}, vol.~71, no.~6, pp. 3752--3762, Apr. 2023.

\bibitem{bellmi}
M.~Bell, ``Information theory and radar waveform design,'' \emph{IEEE Trans.
  Inf. Theory}, vol.~39, no.~5, pp. 1578--1597, Sep. 1993.

\bibitem{tang2018spectrally}
B.~Tang and J.~Li, ``Spectrally constrained {MIMO} radar waveform design based
  on mutual information,'' \emph{IEEE Trans. Signal Process.}, vol.~67, no.~3,
  pp. 821--834, Dec. 2019.

\bibitem{yang2007mimo}
Y.~Yang and R.~S. Blum, ``{MIMO} radar waveform design based on mutual
  information and minimum mean-square error estimation,'' \emph{{IEEE} Trans.
  Aerosp. Electron. Syst.}, vol.~43, no.~1, pp. 330--343, 2007.

\bibitem{chen2013adapt}
Y.~Chen, Y.~Nijsure, C.~Yuen, Y.~H. Chew, Z.~Ding, and S.~Boussakta, ``Adaptive
  distributed {MIMO} radar waveform optimization based on mutual information,''
  \emph{{IEEE} Trans. Aerosp. Electron. Syst.}, vol.~49, no.~2, pp. 1374--1385,
  2013.

\bibitem{wei2024wave}
Z.~Wei, J.~Piao, X.~Yuan, H.~Wu, J.~A. Zhang, Z.~Feng, L.~Wang, and P.~Zhang,
  ``Waveform design for {MIMO-OFDM} integrated sensing and communication
  system: An information theoretical approach,'' \emph{IEEE Trans. Commun.},
  vol.~72, no.~1, pp. 496--509, Jan. 2024.

\bibitem{dong2023rethink}
F.~Dong, F.~Liu, S.~Lu, and Y.~Xiong, ``Rethinking estimation rate for wireless
  sensing: A rate-distortion perspective,'' \emph{{IEEE} Trans. Veh. Technol.},
  vol.~72, no.~12, pp. 16\,876--16\,881, Jul. 2023.

\bibitem{li2023mutual}
J.~Li, G.~Zhou, T.~Gong, and N.~Liu, ``A framework for mutual information-based
  {MIMO} integrated sensing and communication beamforming design,''
  \emph{{IEEE} Trans. Veh. Technol.}, pp. 1--15, 2024.

\bibitem{lijian2007}
J.~Li and P.~Stoica, ``{MIMO} radar with colocated antennas,'' \emph{IEEE
  Signal. Process. Mag.}, vol.~24, no.~5, pp. 106--114, Sept.2007.

\bibitem{Sarwate1980cross}
D.~Sarwate and M.~Pursley, ``Crosscorrelation properties of pseudorandom and
  related sequences,'' \emph{Proc. {IEEE}}, vol.~68, no.~5, pp. 593--619, May
  1980.

\bibitem{robust2013fritz}
R.~Fritzsche and G.~P. Fettweis, ``Robust sum rate maximization in the
  multi-cell {MU-MIMO} downlink,'' in \emph{2013 IEEE Wireless Communications
  and Networking Conference (WCNC)}, Jul. 2013, pp. 3180--3184.

\bibitem{liu2018toward}
F.~Liu, L.~Zhou, C.~Masouros, A.~Li, W.~Luo, and A.~Petropulu, ``Toward
  dual-functional radar-communication systems: Optimal waveform design,''
  \emph{IEEE Trans. Signal Process.}, vol.~66, no.~16, pp. 4264--4279, Jun.
  2018.

\bibitem{Moura2008time}
J.~M.~F. Moura and Y.~Jin, ``Time reversal imaging by adaptive interference
  canceling,'' \emph{IEEE Trans. Signal Process.}, vol.~56, no.~1, pp.
  233--247, Dec. 2008.

\bibitem{2018cheng}
Z.~Cheng, Z.~He, B.~Liao, and M.~Fang, ``{MIMO} radar waveform design with
  {PAPR} and similarity constraints,'' \emph{IEEE Trans. Signal Process.},
  vol.~66, no.~4, pp. 968--981, Feb, 2018.

\bibitem{lehmann2007evaluation}
N.~H. Lehmann, E.~Fishler, A.~M. Haimovich, R.~S. Blum, D.~Chizhik, L.~J.
  Cimini, and R.~A. Valenzuela, ``Evaluation of transmit diversity in
  {MIMO}-radar direction finding,'' \emph{IEEE Trans. Signal Process.},
  vol.~55, no.~5, pp. 2215--2225, Apr. 2007.

\bibitem{hua2023joint}
M.~Hua, Q.~Wu, C.~He, S.~Ma, and W.~Chen, ``Joint active and passive
  beamforming design for irs-aided radar-communication,'' \emph{{IEEE} Trans.
  Wirel. Commun.}, vol.~22, no.~4, pp. 2278--2294, Apr. 2023.

\bibitem{liuAdaptive}
Y.~Liu, G.~Liao, J.~Xu, Z.~Yang, and Y.~Zhang, ``Adaptive {OFDM} integrated
  radar and communications waveform design based on information theory,''
  \emph{IEEE Commun. Lett.}, vol.~21, no.~10, pp. 2174--2177, Jul. 2017.

\bibitem{tang2010mimo}
B.~Tang, J.~Tang, and Y.~Peng, ``{MIMO} radar waveform design in colored noise
  based on information theory,'' \emph{IEEE Trans. Signal Process.}, vol.~58,
  no.~9, pp. 4684--4697, May 2010.

\bibitem{Stoica2007}
P.~Stoica, J.~Li, and Y.~Xie, ``On probing signal design for {MIMO} radar,''
  \emph{IEEE Trans. Signal Process.}, vol.~55, no.~8, pp. 4151--4161, Aug.
  2007.

\bibitem{meyer2000matrix}
C.~D. Meyer, \emph{Matrix analysis and applied linear algebra}.\hskip 1em plus
  0.5em minus 0.4em\relax Siam, 2000, vol.~71.

\bibitem{bjornson2012Pareto}
E.~Björnson, M.~Bengtsson, and B.~Ottersten, ``Pareto characterization of the
  multicell {MIMO} performance region with simple receivers,'' \emph{IEEE
  Trans. Signal Process.}, vol.~60, no.~8, pp. 4464--4469, ,Ay 2012.

\bibitem{bjornson2012robust}
E.~Björnson, G.~Zheng, M.~Bengtsson, and B.~Ottersten, ``Robust monotonic
  optimization framework for multicell {MISO} systems,'' \emph{IEEE Trans.
  Signal Process.}, vol.~60, no.~5, pp. 2508--2523, Jan. 2012.

\bibitem{ye2011interior}
Y.~Ye, \emph{Interior point algorithms: theory and analysis}.\hskip 1em plus
  0.5em minus 0.4em\relax John Wiley \& Sons, 2011.

\bibitem{sen2022frequency}
R.~Senanayake, P.~J. Smith, T.~Han, J.~Evans, W.~Moran, and R.~Evans,
  ``Frequency permutations for joint radar and communications,'' \emph{IEEE
  Trans. Wirel. Commun.}, vol.~21, no.~11, pp. 9025--9040, May 2022.

\bibitem{wen2011sum}
C.-K. Wen, S.~Jin, and K.-K. Wong, ``On the sum-rate of multiuser {MIMO} uplink
  channels with jointly-correlated rician fading,'' \emph{IEEE Trans. Commun.},
  vol.~59, no.~10, pp. 2883--2895, Aug. 2011.

\bibitem{rudin1976principles}
W.~Rudin \emph{et~al.}, \emph{Principles of mathematical analysis}.\hskip 1em
  plus 0.5em minus 0.4em\relax McGraw-hill New York, 1976, vol.~3.

\bibitem{stewart1998matrix}
G.~W. Stewart, \emph{Matrix algorithms: volume 1: basic decompositions}.\hskip
  1em plus 0.5em minus 0.4em\relax SIAM, 1998.

\end{thebibliography}
\end{document}